\documentclass{aastex}          %% The manuscript based on AASTeX v5.x
\usepackage{spr-astr-addons}    %% mimicing ASTR journal style
\begin{document}
%% Article title
%
\title{The oEA stars QY~Aql, BW~Del, TZ~Dra, BO~Her and RR~Lep: Photometric analysis, frequency search and evolutionary status}

%% Running heads
\shorttitle{The oEA stars QY~Aql, BW~Del, TZ~Dra, BO~Her and RR~Lep}

\shortauthors{A. Liakos \& P. Niarchos}

%% Author and Affilations
\author{A. Liakos\altaffilmark{}}
\email{alliakos@phys.uoa.gr} %% non-output
\and
\author{P. Niarchos\altaffilmark{}}
\affil{Department of Astrophysics, Astronomy and Mechanics, National and Kapodistrian University of Athens, GR 157 84, Zografos, Athens, Hellas}

%% Alternate Affilations
%%\altaffiltext{1}{Affilation}
%\altaffiltext{2}{}
%\altaffiltext{3}{}

%% Abstract
\begin{abstract}
New and complete multi-band light cur-ves of the oEA stars QY~Aql, BW~Del, TZ~Dra, BO~Her and RR~Lep were obtained and analysed with the Wilson-Devinney code. The light curves residuals were further analysed with the Fourier method in order to derive the pulsation characteristics of the oscillating components. All the reliable observed times of minimum light were used to examine orbital period irregularities. The orbital period analyses revealed secular changes for QY~Aql and BW~Del, while the Light-Time Effect seems to be the best explanation for the cyclic period changes in TZ~Dra and BO~Her. RR~Lep has a rather steady orbital period. Light curve solutions provided the means to calculate the absolute parameters of the components of the systems, which subsequently were used to make an estimate of their present evolutionary status.
\end{abstract}

%% Keywords
\keywords{Methods: data analysis -- Methods: observational -- stars: binaries:eclipsing -- stars: fundamental parameters -- stars: variables: $\delta$ Scuti -- Stars: evolution -- stars:individual: QY~Aql, BW~Del, TZ~Dra, BO~Her, RR~Lep}

\section{Introduction}
\label{INTRO}

Generally, eclipsing binary systems (hereafter EBs) offer unique information for the calculation of stellar absolute parameters and evolutionary status. Especially, the cases of binaries with $\delta$~Scuti components are extremely interesting, since they provide additional information (i.e. pulsation characteristics) for this part of the stellar lifetime. It has been shown that the $\delta$~Scuti stars in classical Algols (oEA stars) show difference in their pulsational characteristics from time to time due to mass gain \citep{MK04,MK07}. Therefore, the calculation of their absolute parameters and the identification of their oscillating characteristics help us to obtain useful conclusions for this `unstable' part of stellar lifetime. \citet{SO06a} and \citet{LI12} found a connection between orbital and pulsation periods in these systems and showed that binarity plays an essential role in the evolution of the components. Moreover, \citet{LI12} reported an empirical relation between evolutionary stage and dominant pulsation frequency of the $\delta$~Scuti stars in binaries, which differs significantly from that for single ones. Seventy four binaries with $\delta$~Scuti components have been discovered so far \citep{LI12}, but their number is increasing with a rapid rate. The present work is the continuation of the survey for candidate EBs with pulsating components \citep{LN09,LN12,LI12}.

According to the Observed$-$Calculated times of minima variations (hereafter O$-$C) analysis, it is feasible to detect which physical mechanisms play a role in the period modulation (e.g. third body existence, mass transfer between the components) of a binary. On the other hand, from the light curve (hereafter LC) analysis it is possible to determine the Roche geometry of the EB (i.e. semi-detached, detached or contact configuration) or detect a third light. The solutions of these analyses are obviously qualitatively and in some cases also quantitatively (e.g. existence of a third body) connected, even though they are based on different methods.

Five eclipsing systems candidate to include a $\delta$~Sct component, namely V345~Cyg, BW~Del, MX~Her, TW~Lac and AQ~Tau, were selected from the lists of \citet{SO06b} in order to check them for any possible pulsational behaviour. The results showed that only BW~Del exhibits pulsations, therefore systematic observations were performed in order to obtain its complete LCs. Moreover, short-periodic pulsations in the system LT~Her were suspected by Dr. Mkrtichian (private communication), based on his unpublished photometric observations, who kindly suggested us the system for further photometric observations. Our preliminary photometric analysis indeed confirmed the oscillating nature of the primary component and the results will be presented in a future work. Finally, the present work presents results for BW~Del and for other four confirmed cases of oEA stars, namely QY~Aql, TZ~Dra, BO~Her and RR~Lep, for which systematic observations were also made. For these five cases of oEA stars, initially, we performed LC analysis in order to determine their geometric and absolute parameters. Subsequently, frequency analysis on their LC residuals was made with the aim to reveal their main pulsational properties. In addition, since all systems, except for RR~Lep, present orbital period modulations, their O$-$C diagrams were also analysed. Finally, combining the derived information from the pre-mentioned analyses we obtained a more comprehensive view of these systems. The motivation for the present work was: (a) the lack of accurate and/or modern observations for these systems, especially in multiple bands, (b) their poor coverage of their LCs, (c) the lack of accurate pulsation characteristics and (d) the lack of interpretation of their orbital period changes.

%\subsection{Individual systems}
\textbf{QY~Aql}: The system has an orbital period of $\sim7.22956^{\rm d}$. The radial velocities of its primary component and the mass function of the system were calculated by \citet{ST46} and recalculated by \citet{LS71} who found $K_1$=36~km/s and $f(m)=0.035$~M$_{\odot}$, respectively. \citet{GM81}, based on the photographic LCs of \citet{WH45,WH48}, published revised photometric elements of the system but they disputed the past results of $K_1$ and $f(m)$, since both members turned out to be extremely massive. The spectral type of the system is F0 \citep[cf.][]{BU04,MA06}. Modern measurements of the system are given by the \textsl{ASAS} project \citep{PO05}, but they contain only a few points which do not cover the whole LCs. Finally, the pulsational behaviour of its primary was reported by \citet{LI12}.

\textbf{BW~Del}: This EB ($P\sim2.42313^{\rm d}$) was generally neglected. The only available measurement concerns its F2 spectral type \citep[cf.][]{HA04,SK10}.

\textbf{TZ~Dra}: The spectral type of this system is A7V \citep{HE60} and its period has a value $P\sim0.86603^{\rm d}$. \citet{RR90} reported for the first time that the period of the system is changing. \citet{RO03} noticed that small light variations, that can be connected either with spot activity or pulsations, occur in the system. A few years later, \citet{RO05}, \citet{MK05} and \citet{MK06} found $\delta$~Sct-type pulsations in the primary component with a pulsation period of $\sim28^{\rm min}$.

\textbf{BO~Her}: The orbital period of this eclipsing pair is $\sim4.27283^{\rm d}$ and its spectral type is A7V \citep{HA84}. The primary's component oscillating nature was reported by \citet{SB07}, who found a dominant pulsation period of 1.7871$^{\rm hr}$.

\textbf{RR~Lep}: The system has a period of $\sim0.91543^{\rm d}$. Photoelectric LCs were given by \citet{BO86}, \citet{AV89}, \citet{VA89} and \citet{SA89}. \citet{SA89} detected a light variation of $\sim45$~min, but they did not interpret it as a possible pulsation. A CCD LC of the system in $V$-filter was published by the \textsl{ASAS} project \citep{PO05}, but it is of low quality (i.e. small number of points, large photometric error and incomplete LC). \citet{DV09}, based on his CCD observations, found that pulsations occur in the system with a dominant frequency of 31.87~c/d ($\sim45$~min). The spectral type of the system has not been defined so far and ranges between A0-A7 in several catalogues and works \citep[cf.][]{PGC52,MA06,FA02,WR03,SS04}.

The absolute parameters of all systems were calculated by \citet{BD80} (except for QY~Aql), based on the photometric parallax method, and \citet{SK90}, who used statistic relations (e.g. mass-radius, mass-luminosity).

%%%%%%%%%%%%%%%%%%%%%%%%%%%%%%%%%%%%%%%%%%%%%%%%%%%%%%%%%%%%%%%%%%%%%%%%%%%%%%%%%%%%%%%%%%%%%%%%%%%%%%%TABLE 1 --- Log of all obsevations
\begin{table}
\centering
\caption{Observations log of all observed systems.}
\label{tab1}
\scalebox{0.78}{
\begin{tabular}{l ccc c ccc}
\tableline 																			
System	 & $m_{\rm min}^{\rm a}$&    $S.T.$     &	$F$	&	$N$	& $hrs$ &$f_{\rm dom}$	&   $Inst$     \\
	     &	      (mag)	        &		        &		&		&    	&	   (c/d)	&		       \\
\tableline																			
QY Aql	 &	11.4	            &	F0$^{\rm b}$&  $BVI$&	36	&	211	&	   10.656	&  $K$\&$At$   \\
V345 Cyg &	11.3	            &	A1$^{\rm b}$&	$B$	&	2	&	9	&	  	 --	    &	$K$	       \\
BW Del	 &	11.4	            &	F2$^{\rm c}$&	$BV$&	18	&	86	&	   25.100	&	$K$\&$At$  \\
TZ Dra	 &	9.6	                &	A7$^{\rm b}$&	$BV$&	6	&	33	&	    50.993	&	$At$       \\
BO Her	 &	10.7	            &	A7$^{\rm d}$&  $BVI$&	25	&	125 &	   13.430	&	$At$	   \\
LT Her	 &	10.7	            &	A2$^{\rm b}$&  $BV$ &	5	&	20  &	   30.521	&	$At$	   \\
MX Her	 &	11.4	            &	F5$^{\rm b}$&	$B$	&	2	&	12	&        --	    &	$K$\&$At$  \\
TW Lac	 &	11.5	            &	A2$^{\rm c}$&	$B$	&	2	&	10	&	     --	    &	$At$	   \\
RR Lep	 &	10.2	            &	A7$^{\rm c}$&  $BV$	&	9	&	30	&	    33.280  &	$At$	   \\
AQ Tau	 &	12.0	            &	A5$^{\rm b}$&	$B$	&	1	&	4.5	&	 	 --	    &	$At$	   \\
\tableline																					
\multicolumn{8}{l}{$^{\rm a}$\citet{WA06}, $^{\rm b}$\citet{MA06}, $^{\rm c}$\citet{SA11},}\\
\multicolumn{8}{l}{$^{\rm d}$\citet{HA84}}											
\end{tabular}}
\end{table}	
%\vspace{0.5cm}		

%%%%%%%%%%%%%%%%%%%%%%%%%%%%%%%%%%%%%%%%%%%%%%%%%%%%%%%%%%%%%%%%%%%%%%%%%%%%%%%%%%%%%%%%%%%%%%%%%%%%%%%%%%%%%%%%%%%%%%%%%%%%%%%%%%%%%%TABLE 2 --- Log of systematic obsevations
\begin{table*}[t]
\centering
\caption{Detailed observations log of systems with a pulsating component.}
\label{tab2}
\scalebox{0.95}{
\begin{tabular}{l cccc cccc}
\tableline																				
System	&	Nights	&	Obs. dates	&   $T.S.$ 	&\multicolumn{3}{c}{Number of points/$sd$} &	   Comparison 	    &	$m_{\rm V}$     \\
\cline{5-7}
	    &	spent	&	         	&   (d)	    &	 $B$	    &	  $V$	    &	$I$	   &	     stars	        &	    (mag)	    \\
\tableline																					
QY Aql	&	   36	&	28/06-15/09	&	  79    &   3255/3.8    &    3136/3.4	& 3159/3.2 &$C$: TYC 1618-1286-1	&	11.3$^{\rm a}$  \\
	    &	   	    &	   of 2011	&	   	    &               &         	    &          &$K$: TYC 1618-0790-1	&	11.0$^{\rm b}$	\\
BW Del	&	   18	&	01/09-26/10	&	  55    &	1791/3.8    &   1760/4.5	&	--	   &$C$: TYC 1635-1273-1	&	11.4$^{\rm a}$	\\
    	&	      	&	  of 2011	&	  	    &	            &	            &	       &$K$: TYC 1635-1027-2	&	11.4$^{\rm a}$	\\
TZ Dra	&	   6	&	02/07-20/07	&    19	    &	2108/4.3    &   2107/3.5	&	--	   &$C$: TYC 3529-0198-1	&	9.51$^{\rm a}$	\\
    	&	     	&	 of 2008	&	   	    &	            &	            &	       &$K$: TYC 3529-0039-1	&	11.6$^{\rm a}$	\\
BO Her	&	   25	&	28/05-06/07	&	   39   &	1992/3.8    &   1920/3.5	&1881/3.3  &$C$: TYC 2111-0124-1	&	11.3$^{\rm a}$	\\
    	&	   	    &	of 2011	    &	   	    &	            &              	&          &$K$: TYC 2111-0128-1	&	12.3$^{\rm a}$	\\
RR Lep	&	   9	&	17/01-16/03	&	  59	&	1035/2.8    &      991/2.7	&	 --	   &$C$: TYC 5342-0022-1	&	9.6$^{\rm a}$	\\
        &	      	&	 of 2012    &	  	    &	            &            	&		   &$K$: TYC 5342-0128-1	&	10.4$^{\rm a}$	\\
\tableline																				
\multicolumn{9}{l}{$^{\rm a}$\citet{HO00}, $^{\rm b}$\citet{HO98}}									
\end{tabular}}
\end{table*}

\section{Observations and data reduction}
\label{OBS}

The observations were carried out at the Gerostatho-poulion Observatory of the University of Athens ($At$), using the 0.4~m Cassegrain telescope equipped with various CCD cameras and the $BVRI$ \citep[Bessell specification;][]{BE90} photometric filters. In particular, all systems were observed with the ST-10XME CCD, except for TZ~Dra for which the ST-8XMEI CCD was used. Additional observations were also made at the Kryonerion Astronomical Station ($K$) of the Astronomical Institute of the National Observatory of Athens located at Mt.~Kyllini, Corinthia with the 1.2~m Cassegrain telescope equipped with the Ap47p CCD and the Bessell $BVRI$ filters set.

Aperture photometry was applied to the raw data and differential magnitudes were obtained using the software \emph{MuniWin} v.1.1.29 \citep{HR98}. For the cases of BW~Del, TZ~Dra and BO~Her, where field stars exist close to the variables, we chose the photometry apertures with high caution in order to avoid any photometric contribution to the background measurements. The adopted observational strategy in this survey regarding the candidate oscillating binaries was the same as that described in detail in the first paper \citep{LN09}. Briefly, the observational guidelines were: i) the time span should be greater than 3~hr, ii) the filter $B$ and/or $V$ should be used, iii) the comparison star should be of similar magnitude and spectral type to the variable, iv) appropriate exposure times and binning modes must be used for the highest possible photometric $S/N$ (signal-to-noise ratio), and v) the observations should be made outside the primary eclipse.

The log of observations for all systems is listed in Table~\ref{tab1} which contains: their brightest apparent magnitude $m_{\rm min}$, their spectral type $S.T.$, the filters $F$ used, the number of nights $N$ and the total hours $hrs$ spent, the dominant pulsation frequency $f_{\rm dom}$ found (for details see Section~\ref{PULS}) and the instrumentation used $Inst$. Table~\ref{tab2} includes information only for the systems which were observed systematically and pulsations were detected. Particularly, we list: the number of nights spent, the date range and the time span $T.S.$ of the observations, the number of points collected per filter and their mean photometric error $sd$, and the comparison $C$ and check stars $K$ used in the photometry.

\section{Light curve analysis}
\label{DA}

Complete LCs of each system were analysed simultaneously, using all individual observations, with the \emph{PHOEBE} v.0.29d software \citep{PZ05} that is based on the Wilson-Devinney (W-D) code \citep{WD71,WI79,WI90}. In the absence of spectroscopic mass ratios, the `$q$-search' method \citep[cf.][]{LN12} was applied in modes 2 (detached system), 4 (semi-detached binary with its primary component filling its Roche lobe)  and 5 (conventional semi-detached binary) to find feasible (`photometric') estimates for the mass ratio. This value of $q$ was set as adjustable parameter in the subsequent analysis. The temperatures of the primaries $T_1$ of all systems, except for RR~Lep, were assigned values according to the spectral class-temperature correlation \citep{CO00} and were kept fixed during the analysis, while the temperatures of secondaries $T_2$ were adjusted. The values of bolometric albedos $A$ and gravity darkening coefficients $g$ were given standard theoretical values according to the adopted type of stellar atmosphere, namely $A$=1 and $g$=1 for radiative \citep{RU69,VZ24}, and $A$=0.5 and $g$=0.32 for convective atmospheres \citep{RU69,LU67}. The (linear) limb darkening coefficients, $x_1$ and $x_2$, were taken from \citet{VH93}; the dimensionless potentials $\Omega_{1}$ and $\Omega_{2}$, the fractional luminosity of the primary component $L_{1}$ and the system's orbital inclination $i$ were set as adjustable parameters.

%%%%%%%%%%%%%%%%%%%%%%%%%%%%%%%%%%%%%%%%%%%%%%%%%%%%%%%%%%%%%%%%%%%%%%%%%%%%%%%%%%%%%%%%%%%%%%%%%%%%%% Figure 1 -- LCs
\begin{figure}
\centering
\begin{tabular}{cl}
%QY Aql&BW Del\\
\includegraphics[width=7.1cm]{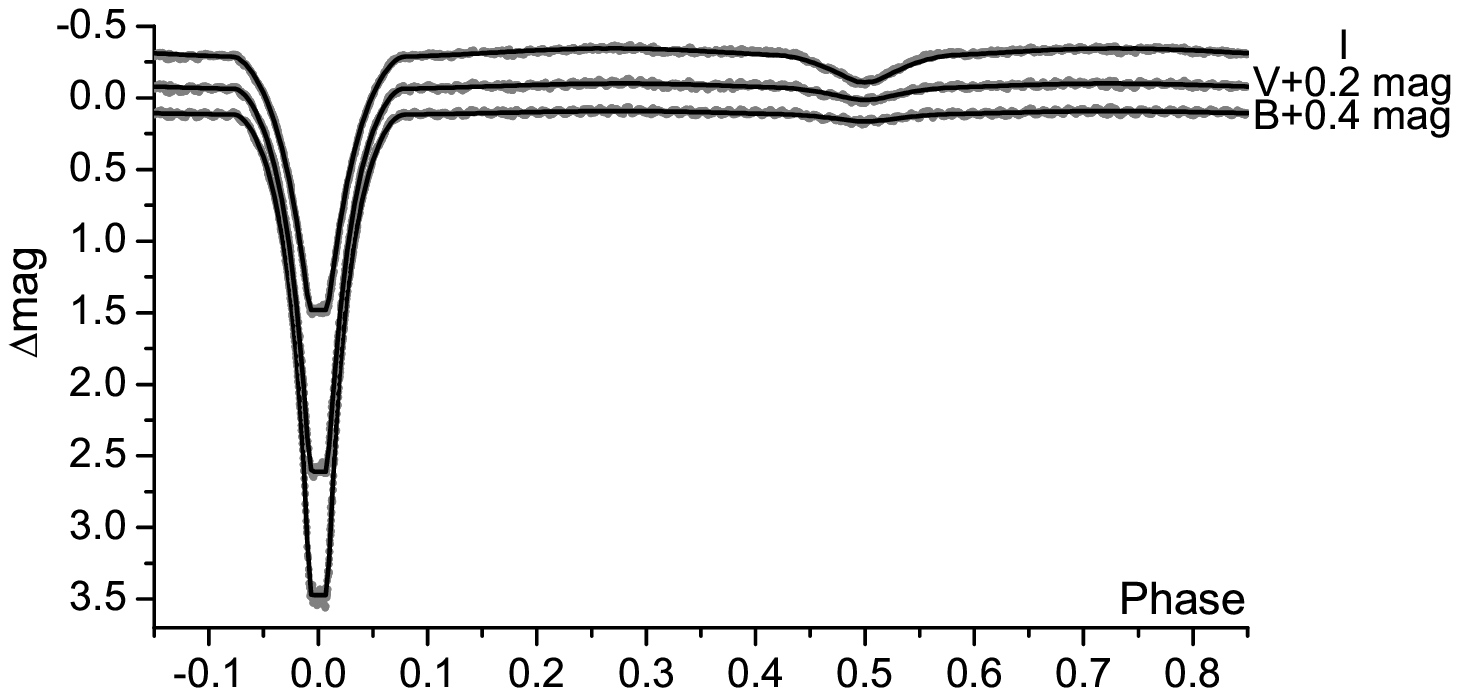}&(a)\\
\includegraphics[width=7.1cm]{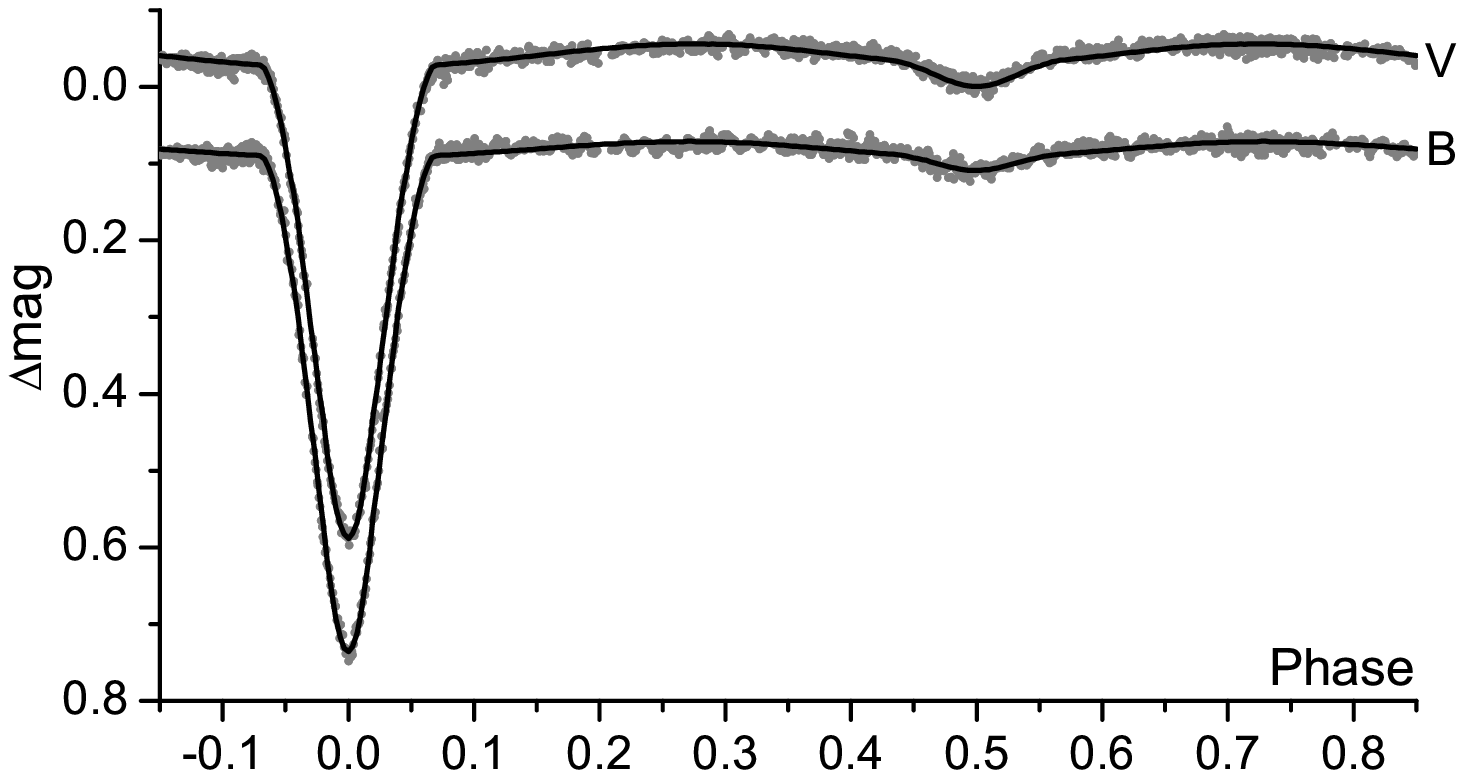}&(b)\\
\includegraphics[width=7.1cm]{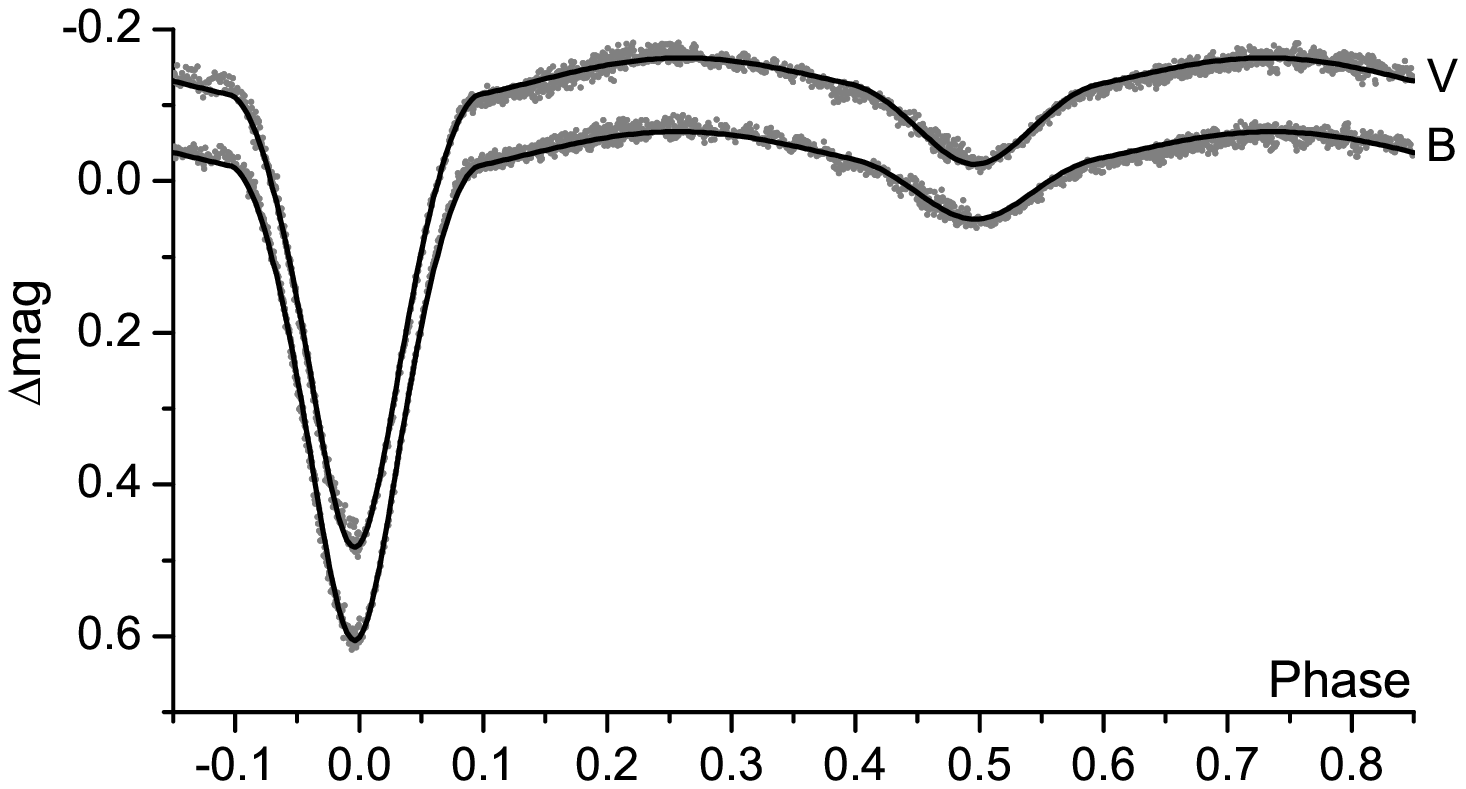}&(c)\\
%TZ Dra&BO Her\\
\end{tabular}
\caption{Synthetic (solid lines) and observed (points) light curves of (a) QY~Aql, (b) BW~Del, (c) TZ~Dra.}
\label{fig1}
\end{figure}

%%%%%%%%%%%%%%%%%%%%%%%%%%%%%%%%%%%%%%%%%%%%%%%%%%%%%%%%%%%%%%%%%%%%%%%%%%%%%%%%%%%%%%%%%%%%%%%%%%%%%% Figure 2 -- LCs
\begin{figure}[t]
\centering
\begin{tabular}{cl}
%TZ Dra&BO Her\\

\includegraphics[width=7.1cm]{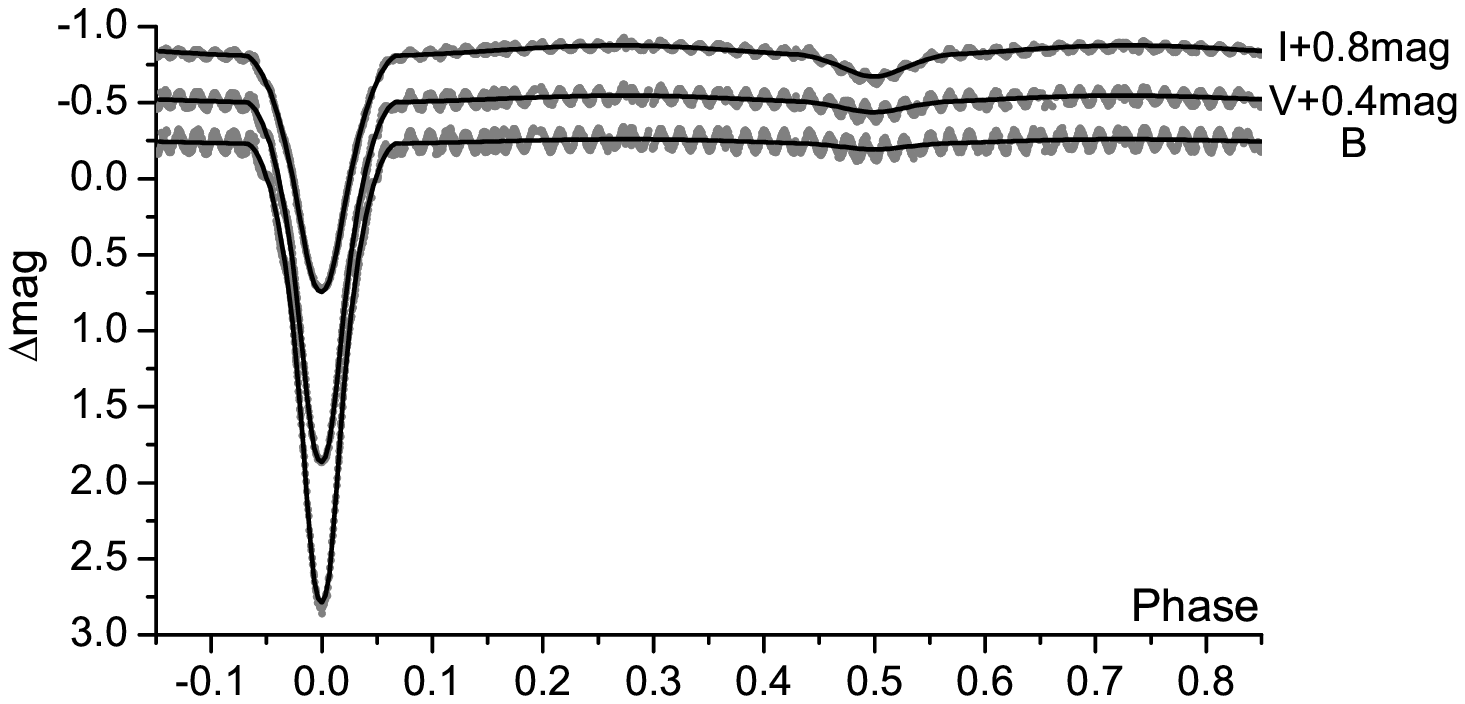}&(a)\\
%RR Lep&\\
\includegraphics[width=7.1cm]{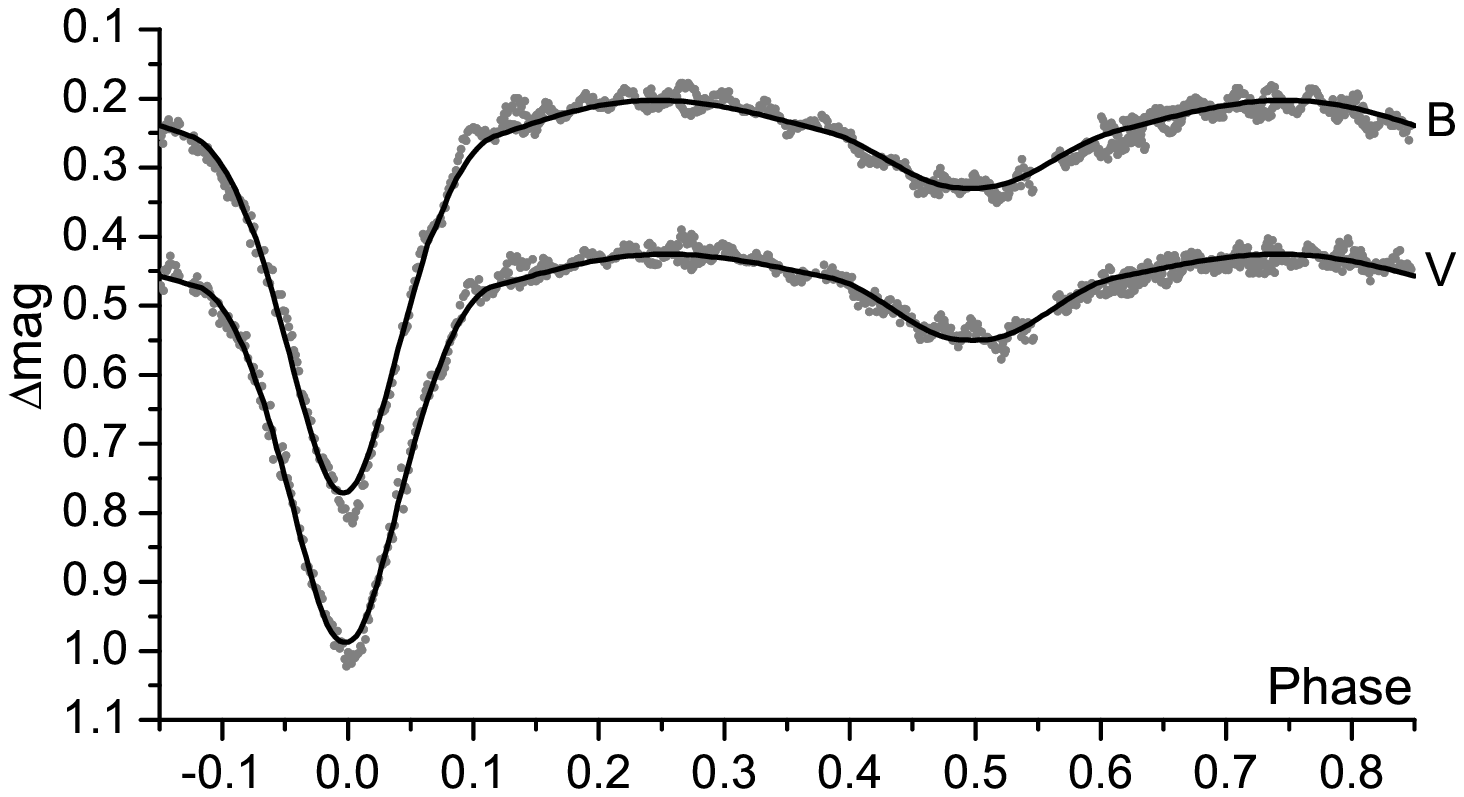}&(b)\\
\end{tabular}
\label{fig2}
\caption{Synthetic (solid lines) and observed (points) light curves of (a) BO~Her and (b) RR~Lep.}
\end{figure}

%%%%%%%%%%%%%%%%%%%%%%%%%%%%%%%%%%%%%%%%%%%%%%%%%%%%%%%%%%%%%%%%%%%%%%%%%%%%%%%%%%%%%%%%%%%%%%%%%%%%%%%TABLE 3 ---- LCs parameters
\begin{table*}
\centering
\caption{Light curve (upper part) and absolute parameters (lower part) for all systems. Formal errors are indicated in parentheses alongside adopted values.}
\label{tab3}
\scalebox{0.82}{
\begin{tabular}{l cc cc cc cc cc}
\tableline																					
                                                                                        \multicolumn{11}{c}{Light curve parameters}	                                                          \\
\tableline																					
Parameter	                 &	\multicolumn{2}{c}{QY Aql}	  &	 \multicolumn{2}{c}{BW Del}	 &	\multicolumn{2}{c}{TZ Dra}	 &	\multicolumn{2}{c}{BO Her}	  &	 \multicolumn{2}{c}{RR Lep}	  \\
\tableline																					
$i~(^\circ$)	             &  \multicolumn{2}{c}{88.6 (5)}  &	\multicolumn{2}{c}{78.6 (4)} &	\multicolumn{2}{c}{77.6 (1)} &	\multicolumn{2}{c}{85.4 (4)}  &	 \multicolumn{2}{c}{80.3 (9)}  \\
$q~ (m_{2}/m_{1}$)	         & \multicolumn{2}{c}{0.25 (2)}   &	\multicolumn{2}{c}{0.16 (2)} &	\multicolumn{2}{c}{0.31 (3)} &	\multicolumn{2}{c}{0.22 (2)}  &	 \multicolumn{2}{c}{0.23 (2)}  \\
\tableline																					
$Component$	                 &   	  P	      &	       S	  &	      P       &	     S	     &	       P	  &	       S	 &	      P	      &	      S	      &	      P	      &	      S	      \\
\tableline																					
$T$ (K)                      &	7300$^{\rm a}$&    4244 (122) &	7000$^{\rm a}$&	  4061 (30)	 &	7800$^{\rm a}$&	  5088 (55)	 &	7800$^{\rm a}$&	   4344 (68)  &	 7800$^{\rm a}$&	4925 (96)	  \\
$\Omega$	                 &	5.19 (6)	  &	2.34$^{\rm b}$&	   4.61 (6)	  &2.14$^{\rm b}$&	   3.46 (1)	  &2.49$^{\rm b}$&	     6.24 (1) &	2.29$^{\rm b}$&	    2.74 (2)  &	2.32$^{\rm b}$\\
$x_{\mathrm{B}}$	         &	    0.669	  &	   1.010	  &	     0.592    &	   0.957	 &	   0.595	  &	    0.850	 &	     0.609	  &	    0.990	  &	   0.600	  &	   0.866	  \\
$x_{\mathrm{V}}$	         &	    0.569	  &	   0.856	  &	     0.493    &	   0.823	 &	   0.523	  &	    0.707	 &	     0.529	  &	    0.835	  &	   0.525	  &	   0.746	  \\
$x_{\mathrm{I}}$	         &	    0.385	  &	    0.584	  &	        --	  &	     --	     &	      --	  &	      --	 &	     0.361	  &	  0.581	      &	      --	  &	      --	  \\
$(L/L_\mathrm{T})_\mathrm{B}$&	   0.942 (1)  &	   0.058 (1)  &	     0.963 (1)&	   0.037 (2) &	  0.924 (1)	  &	    0.076 (1)&	     0.923 (6)&	   0.077 (1)  &	     0.964 (3)&	   0.036 (2)  \\
$(L/L_\mathrm{T})_\mathrm{V}$&	   0.896 (1)  &	   0.104 (1)  &	     0.932 (1)&	  0.068 (1)	 &	   0.889 (1)  &	    0.111 (1)&	     0.866 (5)&	   0.134 (1)  &	     0.952 (2)&	   0.048 (2)  \\
$(L/L_\mathrm{T})_\mathrm{I}$&	   0.773 (1)  &	   0.227 (1)  &	       --	  &	       --	 &      	--	  &	      --	 &	    0.730 (3) &	   0.270 (1)  &	     --	      &	     --	      \\
\tableline																				
                                                                                          \multicolumn{11}{c}{Absolute parameters}		                                                      \\				
\tableline																					
$M~(M_{\sun}$)	             &1.6 (2)$^{\rm a}$&	0.4 (1)	  &1.5 (2)$^{\rm a}$&	0.3 (1)	 &1.8 (2)$^{\rm a}$&	0.6 (1)	 &1.8 (2)$^{\rm a}$&	0.4 (1)	  &1.8 (2)$^{\rm a}$&	0.4 (1)	  \\
$R~(R_{\sun}$)	             &	  4.1 (2)	  &	    5.4 (2)	  &	    2.1 (1)	  &	    2.2 (1)	 &	    1.7 (1)	  &	   1.5 (1)	 &	   2.5 (1)	  &	    3.8 (1)	  &	    2.2 (1)	  & 	1.4 (2)	  \\
$L~(L_{\sun}$)	             &	  43 (3)	  &	     8 (1)	  &	   10 (1)	  &	    1.2 (1)	 &	     9 (1)	  &	   1.3 (1)	 & 	   20 (1)	  &	    4.6 (4)	  &	   15.6 (4)	  &	   1.0 (1)	  \\
$\log g$ (cm/s$^2$)	         &	 3.4 (1)	  &	    2.6 (1)	  &	    4.0 (1)	  &	    3.1 (1)	 &	     4.2 (1)  &	   3.9 (1)	 & 	  3.9 (1)	  &	    2.9 (1)	  &	    4.0 (1)	  &	    3.8 (2)	  \\
$a~(R_{\sun}$)	             &	  4.0 (2)	  &	    16.3 (7)  &	    1.3 (1)	  & 	8.0 (4)	 &	    1.2 (1)	  &	    4.0 (2)	 &	 2.7 (1)	  &	    12.1 (5)  &	    1.0 (1)	  &	    4.3 (2)	  \\
\tableline																					
\multicolumn{11}{l}{$^{\rm a}$assumed, $^{\rm b}$fixed, $L_{\mathrm{T}}= L_1+L_2$, P=Primary, S=Secondary}																					
\end{tabular}}
\end{table*}

Since the spectral type of RR~Lep ranges between A0-A7, the above analysis' steps were made for different values of $T_1$ and the final solution was selected according to the least value of squared residuals.

All primaries were adopted as radiative and all secondaries as convective stars according to their temperature values, therefore we set $A_1$=1, $A_2$=0.5, $g_1$=1 and $g_2$=0.32. In the cases of TZ~Dra and BO~Her the relative luminosity contribution $l_{3}$ of a possible third light was left free due to possible existence of tertiary components (see Section~\ref{O-C}). Nevertheless, it resulted in unrealistic values for both systems, therefore it was excluded from the final solutions. Finally, all systems were found to be in semi-detached configurations with their cooler and less massive components filling their Roche lobes. Observed LCs and their modelling are illustrated in Figs~1-2 with corresponding parameters listed in Table~\ref{tab3}.

\section{Absolute parameters and evolutionary status of the components}
\label{ABS}
Although no radial velocity measurements exist for the systems studied, we can form fair estimates of their absolute parameters. Since there is no trustable information in the literature (see Section~\ref{INTRO}, regarding the methods used for the determination of their absolute parameters, the masses of the primaries were assumed according to their spectral types using the correlations of \citep{CO00}. A fair error of $\sim10$\% of the mass value was also assumed in order to obtain more realistic conclusions. The secondary masses follow from the determined mass ratios (see Table~\ref{tab3}) and the semi-major axes $a$ are then derived from Kepler's third law. The errors were calculated using the error propagation method. The parameters are listed in Table~\ref{tab3} and the positions of the systems' components in the $M-R$ diagram are given in Fig.~3. The theoretical lines for Zero Age Main Sequence (ZAMS) and Terminal Age Main Sequence (TAMS) were taken from \citet{NM03}.

The primary of TZ~Dra is located closer to the ZAMS, while the primary of RR~Lep closer to the TAMS. The primary of BO~Her was found to be exactly on the TAMS, while the primaries of BW~Del and QY~Aql have left the MS. All secondaries are evolved stars lying far beyond the TAMS limits.

%%%%%%%%%%%%%%%%%%%%%%%%%%%%%%%%%%%%%%%%%%%%%%%%%%%%%%%%%%%%%%%%%%%%%%%%%%%%%%%%%%%%%%%%%%%%%%%%%%%%%% Figure 3 -- M-R
\begin{figure}[t]
\centering
\includegraphics[width=8cm]{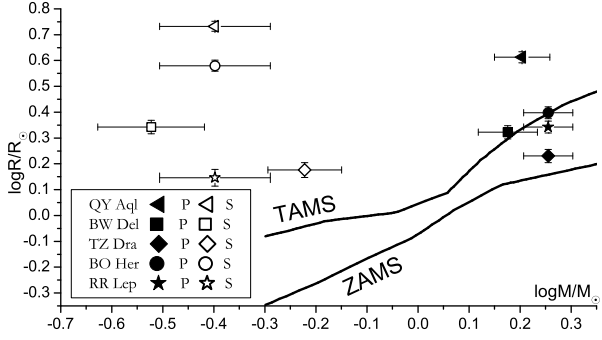}
\label{fig3}
\caption{Positions of the systems' components (\textsl{P}=Primary, \textsl{S}=Secondary) in the $M-R$ diagram.}
\end{figure}

\section{Frequency analysis}
\label{PULS}

The pulsating components of the systems are the primaries, since their temperatures are well inside the range of $\delta$~Sct type stars (A-F spectral types). For the frequency search, the theoretical LCs of the eclipsing binary model were subtracted from the respective observed data. Frequency analysis was performed on the LC residuals on the out of primary eclipse data with the software \emph{PERIOD04} v.1.2 \citep{LB05}, that is based on classical Fourier analysis. Given that typical frequencies for $\delta$~Sct stars range between 3-80~c/d \citep{BR00,SO06b}, the analysis was made for this range. Frequencies in the range 0-3~c/d were considered as non-physical, and they were excluded from the final model. After the first frequency computation the residuals were subsequently pre-whitened for the next one, until the detected frequency had $S/N<4$, which is the programme's critical trustable limit. The errors were calculated using analytical simulations. The $l$-degrees of the pulsation modes were identified with the software \emph{FAMIAS} v.1.01 \citep{ZI08} that is based on theoretical $\delta$~Scuti models \citep[$MAD$~-][]{MD07}. However, the $l$-degrees determination, using only photometric data, is strongly based on the information from various wavelength bands. Since we used only two or three filters for these systems, the $l$-degrees calculations can be considered as preliminary. Frequency analysis results are given in Table~\ref{tab4}, where we list: frequency values $f$, $l$-degrees, semi-amplitudes $A$, phases $\Phi$ and $S/N$. Amplitude spectra, spectral window plots and Fourier fits on the longest data sets are given in Fig.~4.

For QY~Aql one pulsation frequency ($\sim10.656$~c/d) was detected. For its adopted mass value, there is no theoretical $\delta$~Sct model for determining the $l$-degree, therefore, we tested another slightly higher mass values. We found that the $l$-degree can be calculated using a mass of 1.9~$M_{\odot}$.

TZ~Dra is found to oscillate in a mono-periodic mode with a frequency value $\sim50.994$~c/d, while its $l$-degree was calculated using the adopted mass value.

Three oscillation frequencies were found for BW~Del with the most dominant one at $\sim25.100$~c/d. The $l$-degrees were determined using a mass value of 1.6~$M_{\odot}$, which is inside the limits of the adopted error. Two pulsation frequencies were identified for BO~Her and RR~Lep. However, for BO~Her the frequencies $f_2$ and $f_4$ are the first and second harmonics of $f_1$, respectively, while $f_3$ and $f_4$ were found below the significance limit in $I$-filter data and they are excluded from the final solution. The $l$-degrees for both systems were calculated using the mass values given in Table~\ref{tab3}.

%%%%%%%%%%%%%%%%%%%%%%%%%%%%%%%%%%%%%%%%%%%%%%%%%%%%%%%%%%%%%%%%%%%%%%%%%%%%%%%%%%%%%%%%%%%%%%%%%%%%%%%TABLE 4 ----Freqs
\begin{table*}
\caption{Frequency analyses results. The errors are indicated in parentheses alongside adopted values.}
\label{tab4}
\centering
\scalebox{0.79}{
\begin{tabular}{cc cccc cccc cccc}
\tableline 																											
$No$	&	$l$	&	 $f$	&$A$ 	 &	$\Phi$	&	S/N	&	 $f$	&  $A$ 	&	$\Phi$	&	S/N	&	 $f$   &  $A$ 	&	$\Phi$	&	S/N	\\
	    &		&  (c/d)	& (mmag) &($^\circ$)&	    &  (c/d)	&(mmag)	&($^\circ$)	&		&   (c/d)  &(mmag)	&($^\circ$)	&		\\
\tableline																											
	    &		&   \multicolumn{4}{c}{$B$-filter}	    &	    \multicolumn{4}{c}{$V$-filter}	&		\multicolumn{4}{c}{$I$-filter}	\\
\tableline																											
															\multicolumn{14}{c}{QY Aql}											        \\
\tableline																											
$f_1$	&	1	&10.6561 (1)&11.8 (2)&	23 (1)	&35.8	&10.6562 (2)&9.4 (2)&	26 (1)	&	17.7&10.6560 (3)&5.1 (2)&	22 (2)	&	11.5\\
\tableline																											
															\multicolumn{14}{c}{BW Del}												    \\
\tableline																											
$f_1$   &1 or 3	&25.100 (1)	&2.9 (2) &	74 (4)	&5.1	&25.100 (1)	&1.8 (2)&	89 (6)	&	4.7	&	--	    &	--	&	 --	    &	--	\\
$f_2$   &0 or 2	&19.641 (1)	&1.9 (2) &	46 (6)	&4.5	&19.643 (1)	&1.2 (2)&	26 (9)	&	4.4	&	--	    &	--	&	 --	    &	--	\\
$f_3$   &1 or 2	&27.608 (1)	&1.7 (2) &	324 (7)	&4.1	&27.608 (1)	&1.3 (2)&	328 (9)	&	4.5	&	--	    &	--	&	 --	    &	--	\\
\tableline																											
															\multicolumn{14}{c}{TZ Dra}											     	\\
\tableline																											
$f_1$   &1 or 3	&50.993 (2)	&3.7 (2) &	82 (4)	&6.9    &50.995 (3)	&2.8 (2)&	70 (5)	&	7.0	&	  --	&	--	&	  --	&	--	\\
\tableline																											
															\multicolumn{14}{c}{BO Her}												    \\
\tableline																											
$f_1$	&	3	&13.430 (1)	&68.0 (3)&	137 (1)	&173.8	&13.429 (1)	&50.8 (3)&	138 (1)	&	58.8&13.429 (1)	&25.6 (3)&	137 (1)	&	39.8\\
$f_2$	&		&26.860 (1)	&5.6 (3) &	65 (3)	&12.5	&26.860 (1)	&4.5 (3) &	77 (4)	&	10.4&26.859 (2)	&2.4 (3) &	77 (8)	&	6.4	\\
$f_3$   &0 or 1	&23.057 (1)	&3.2 (3) &	227 (5)	&9.0	&23.058 (2)	&2.2 (3) & 	228 (7)	&	5.0	&	  --	&	 --	 &	    --	&	--	\\
$f_4$	&		&40.293 (3)	&1.3 (3) &	231 (11)&5.5	&40.296 (4)	&1.1 (3) &	176 (15)&	4.0	&	  --	&	 --	 &	   --	&	--	\\
\tableline																											
															\multicolumn{14}{c}{RR Lep}												    \\
\tableline																											
$f_1$   &  3	&33.2802 (4)& 9.6 (4)&  86 (2)	& 6.4  &33.2709 (5)	&7.6 (4) &	91 (3)	&	6.1	&	  --	&	 --	 &	  --	&	--	\\
$f_2$   &0 or 1 &24.318 (1)	& 3.5 (4)&  149 (6)	& 4.0  &24.560 (11)	&3.4 (4) &	66 (7)	&	4.0	&	  --	&	 --	 &	  --	&	--	\\
\tableline																											
\end{tabular}}
\end{table*}

%%%%%%%%%%%%%%%%%%%%%%%%%%%%%%%%%%%%%%%%%%%%%%%%%%%%%%%%%%%%%%%%%%%%%%%%%%%%%%%%%%%%%%%%%%%%%%%%%%%%%% Figure 4 -- AS & FF
\begin{figure*}
\centering
\begin{tabular}{ccl}
\includegraphics[width=7.5cm]{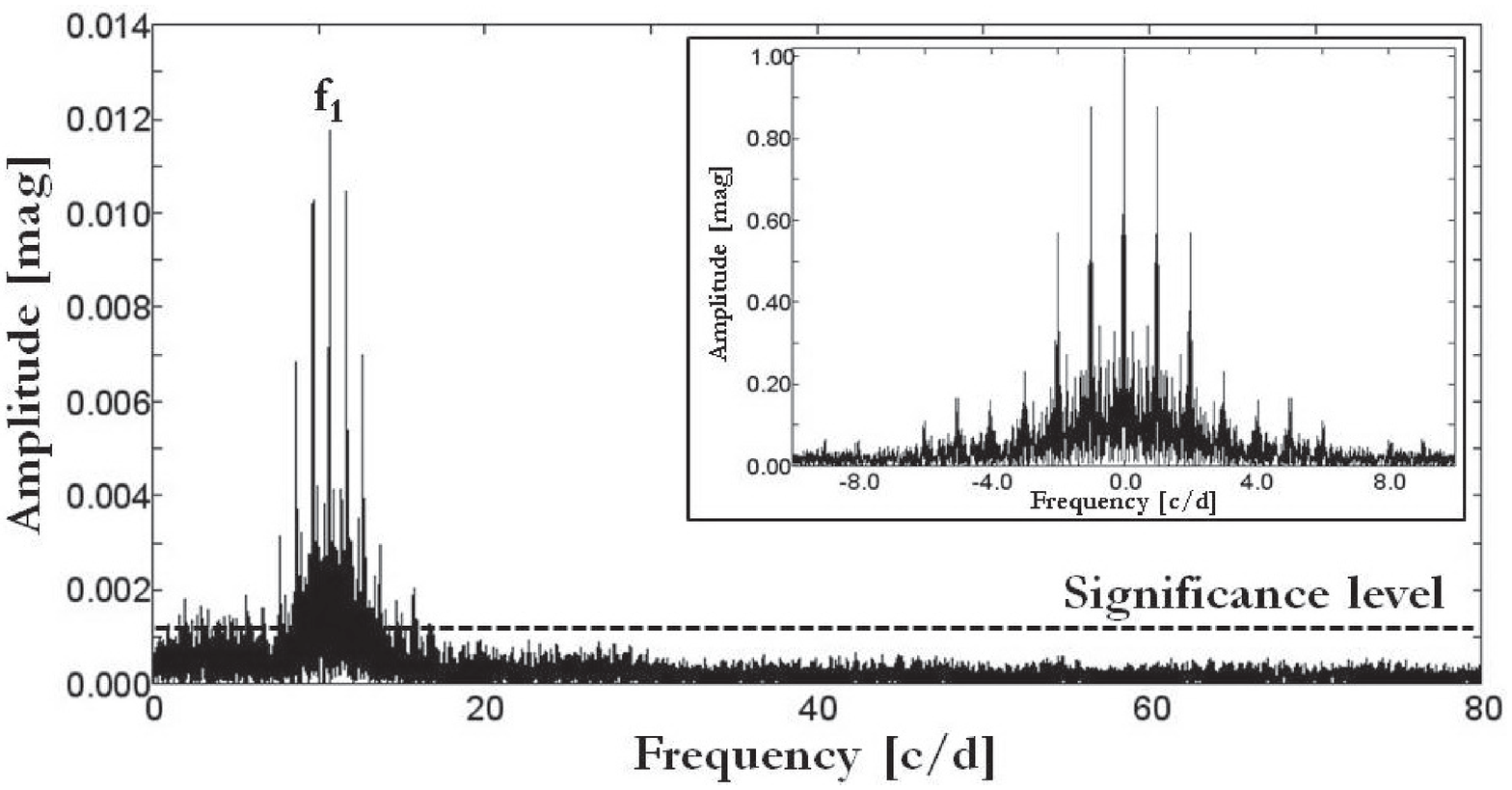}&\includegraphics[width=7.5cm]{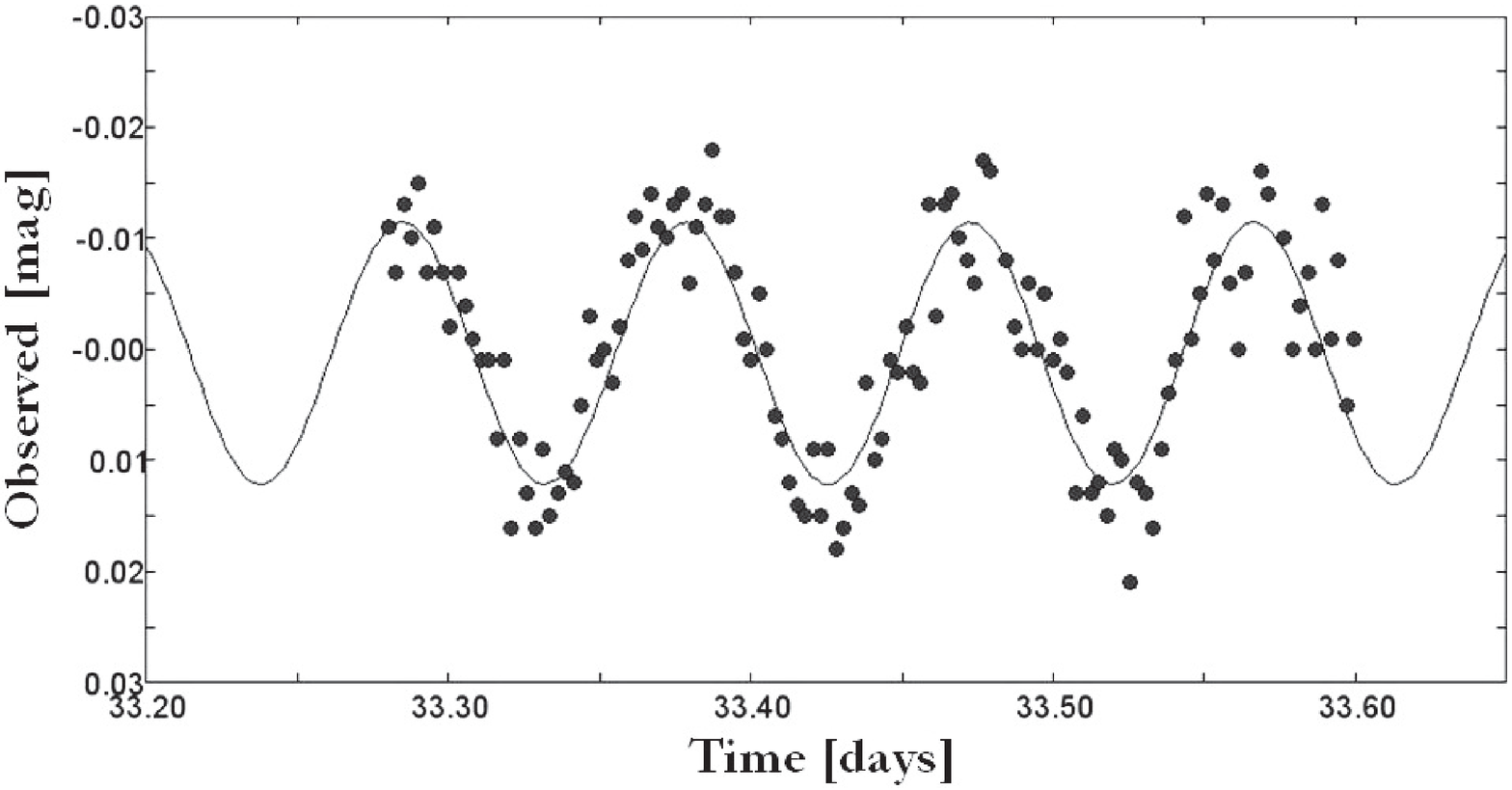}&(a)\\
\includegraphics[width=7.5cm]{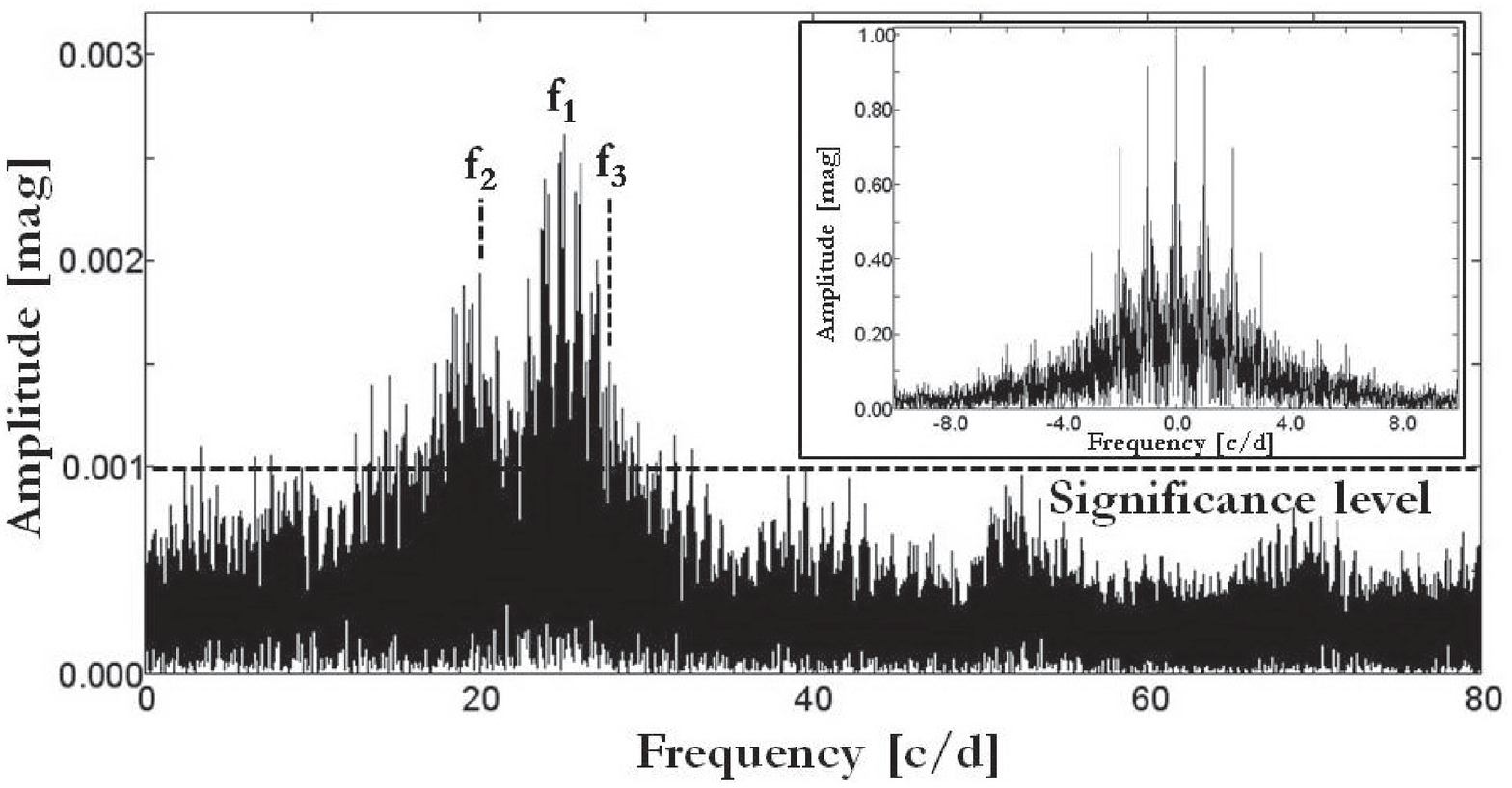}&\includegraphics[width=7.5cm]{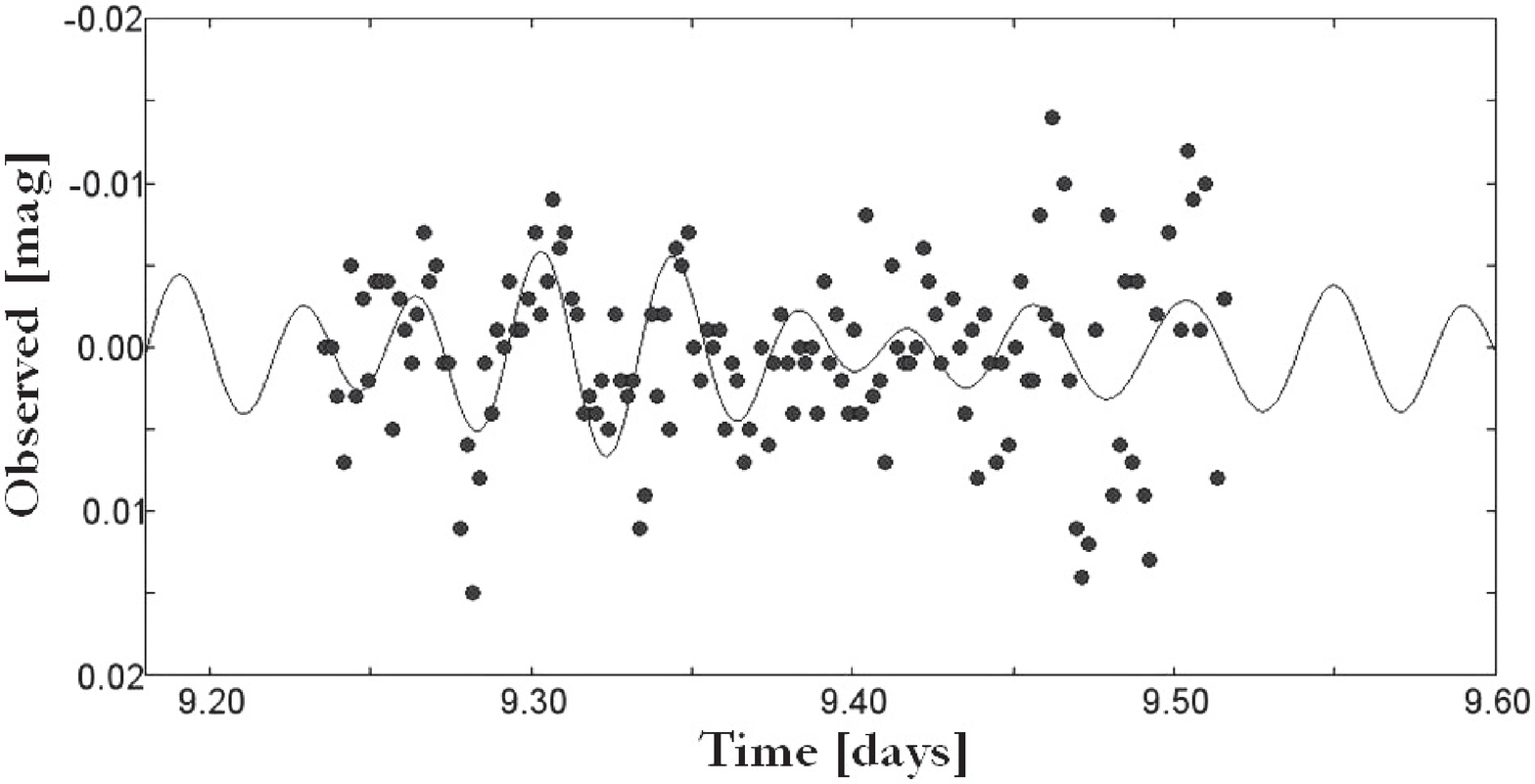}&(b)\\
\includegraphics[width=7.5cm]{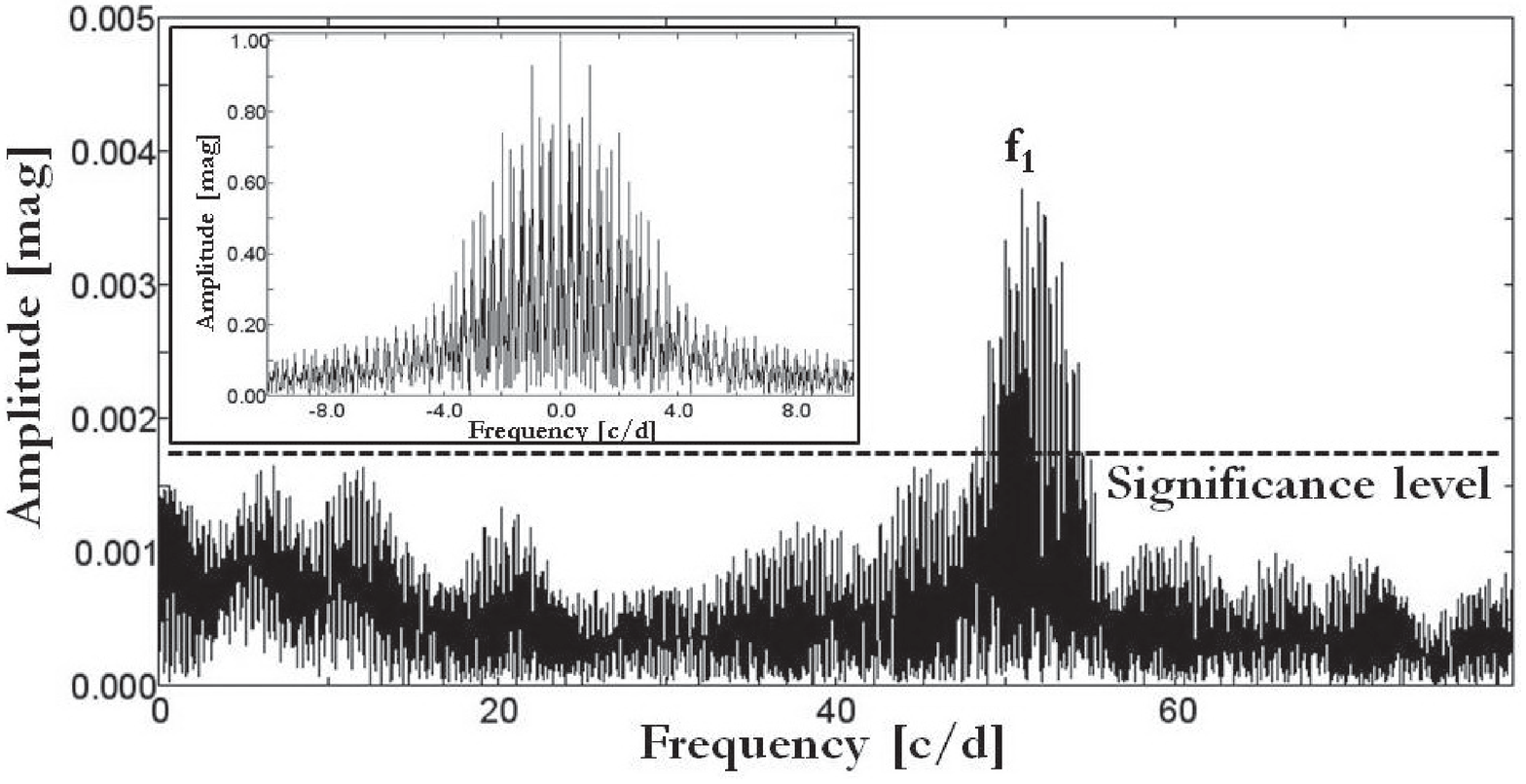}&\includegraphics[width=7.5cm]{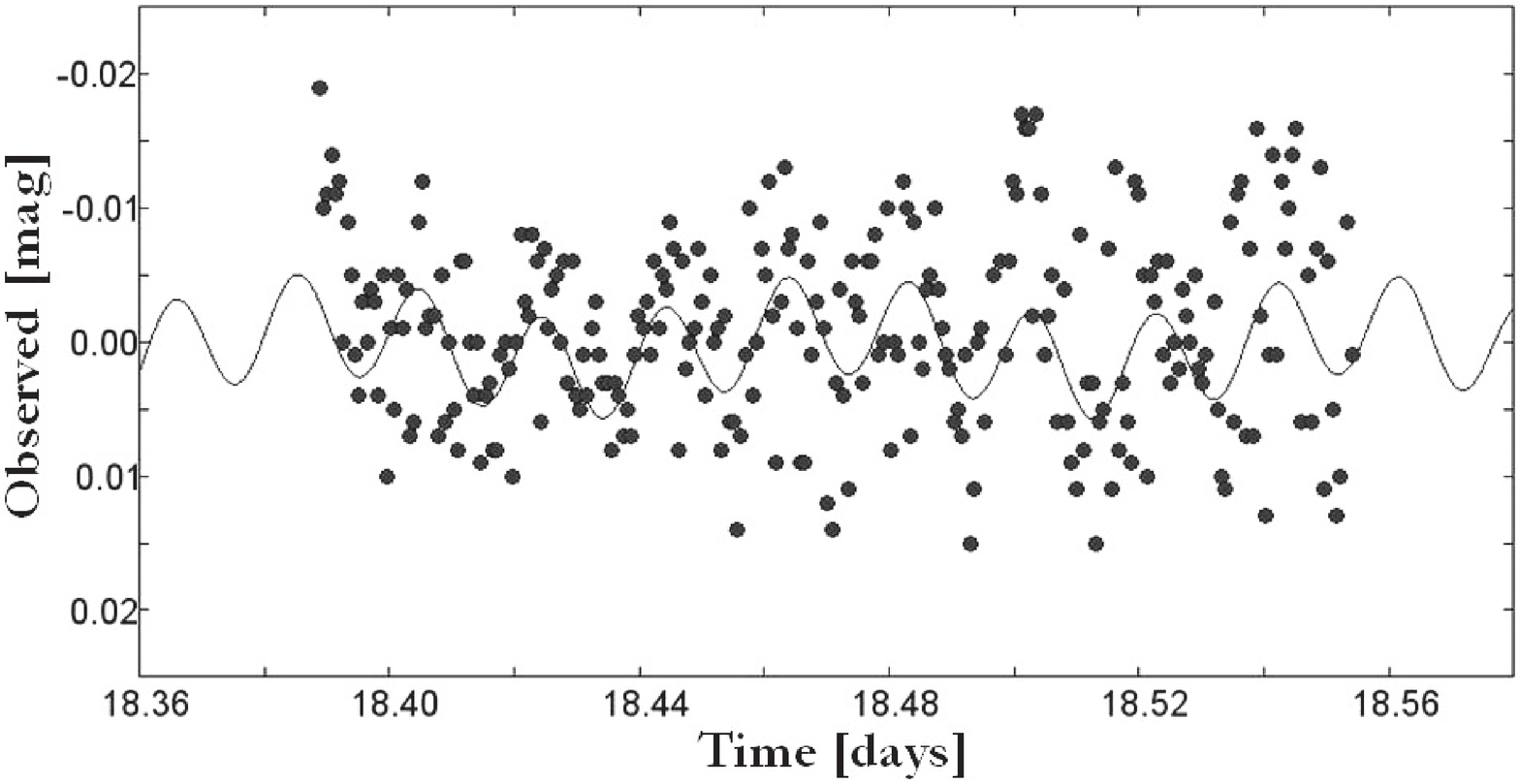}&(c)\\
\includegraphics[width=7.5cm]{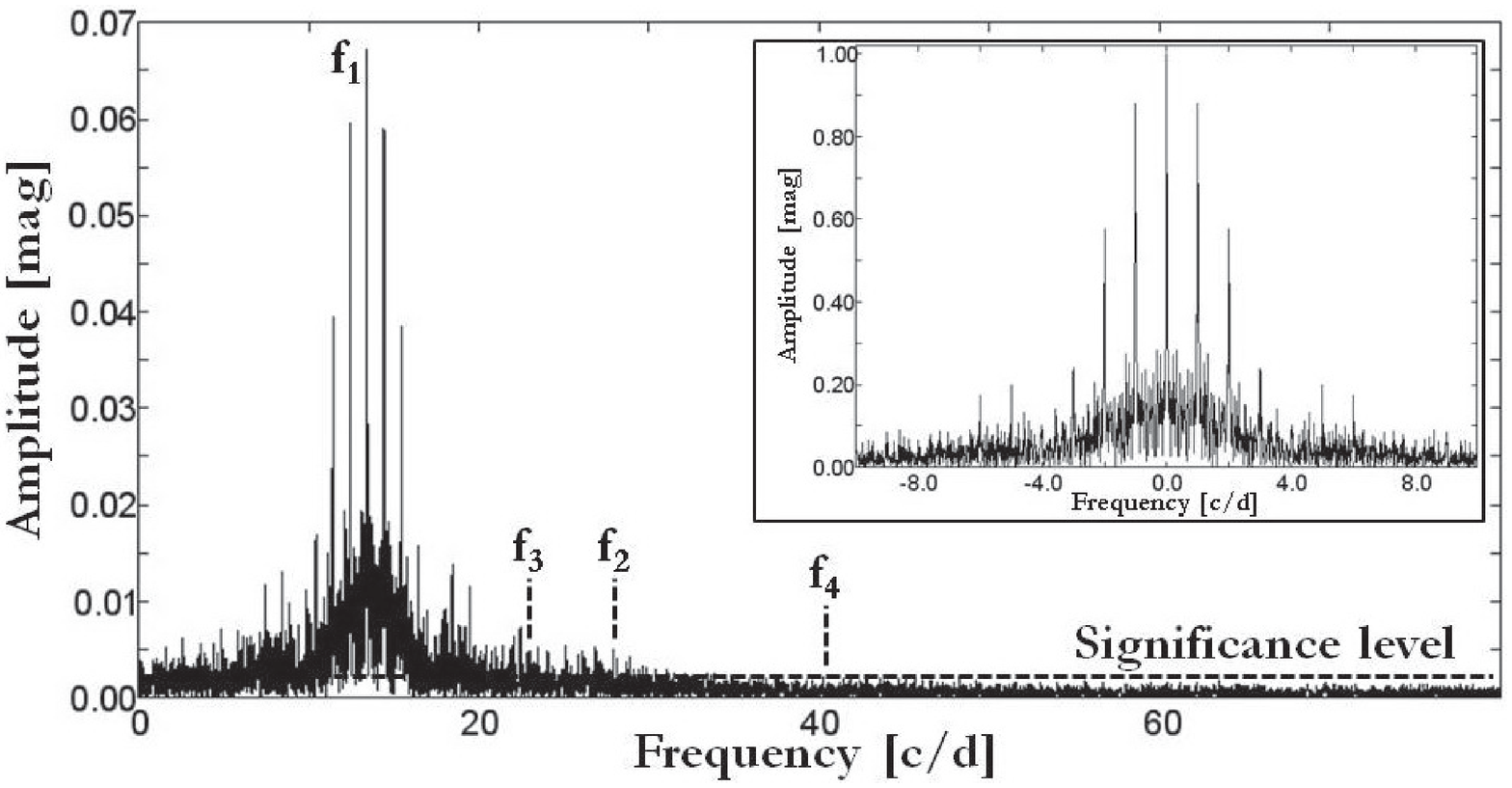}&\includegraphics[width=7.5cm]{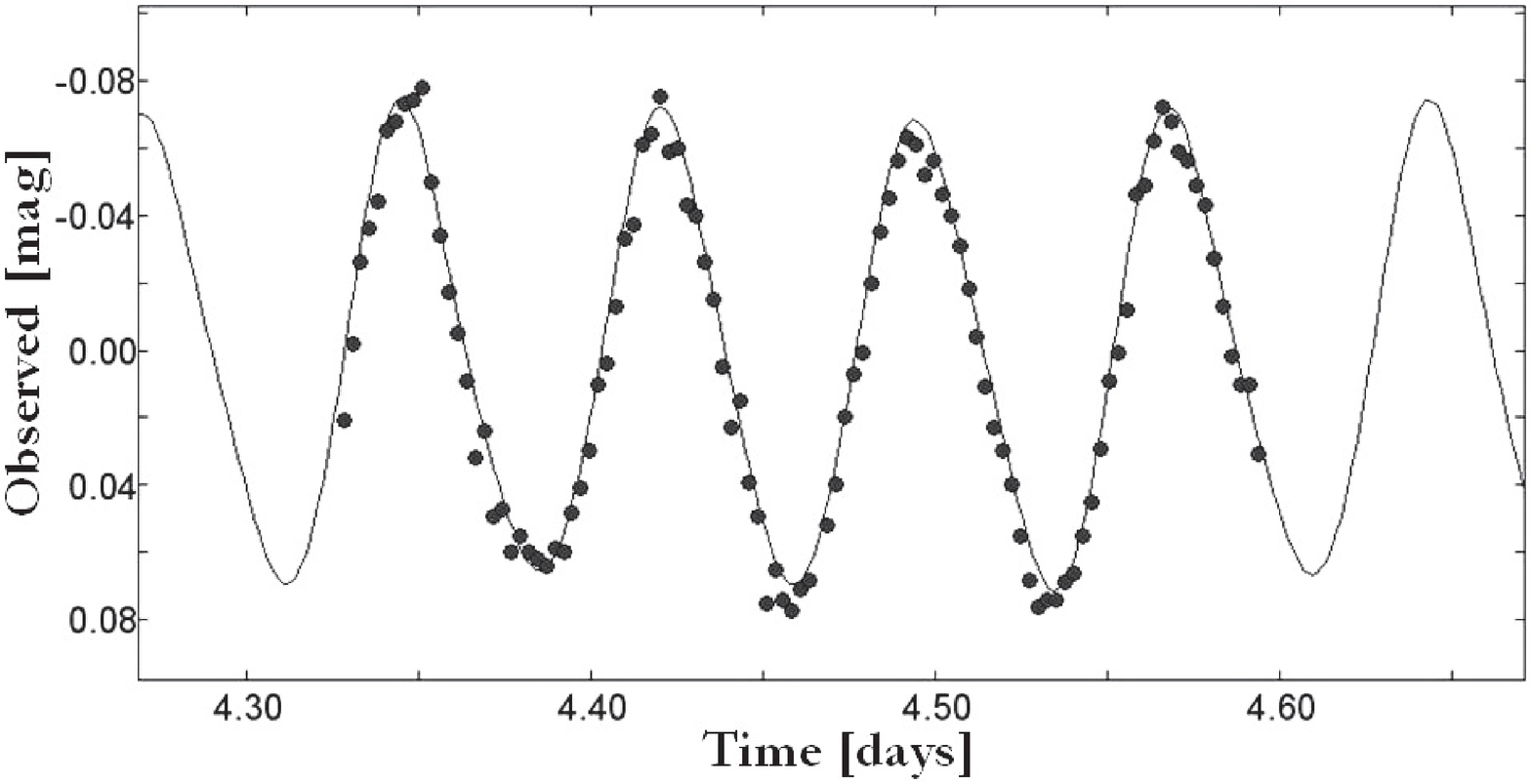}&(d)\\
\includegraphics[width=7.5cm]{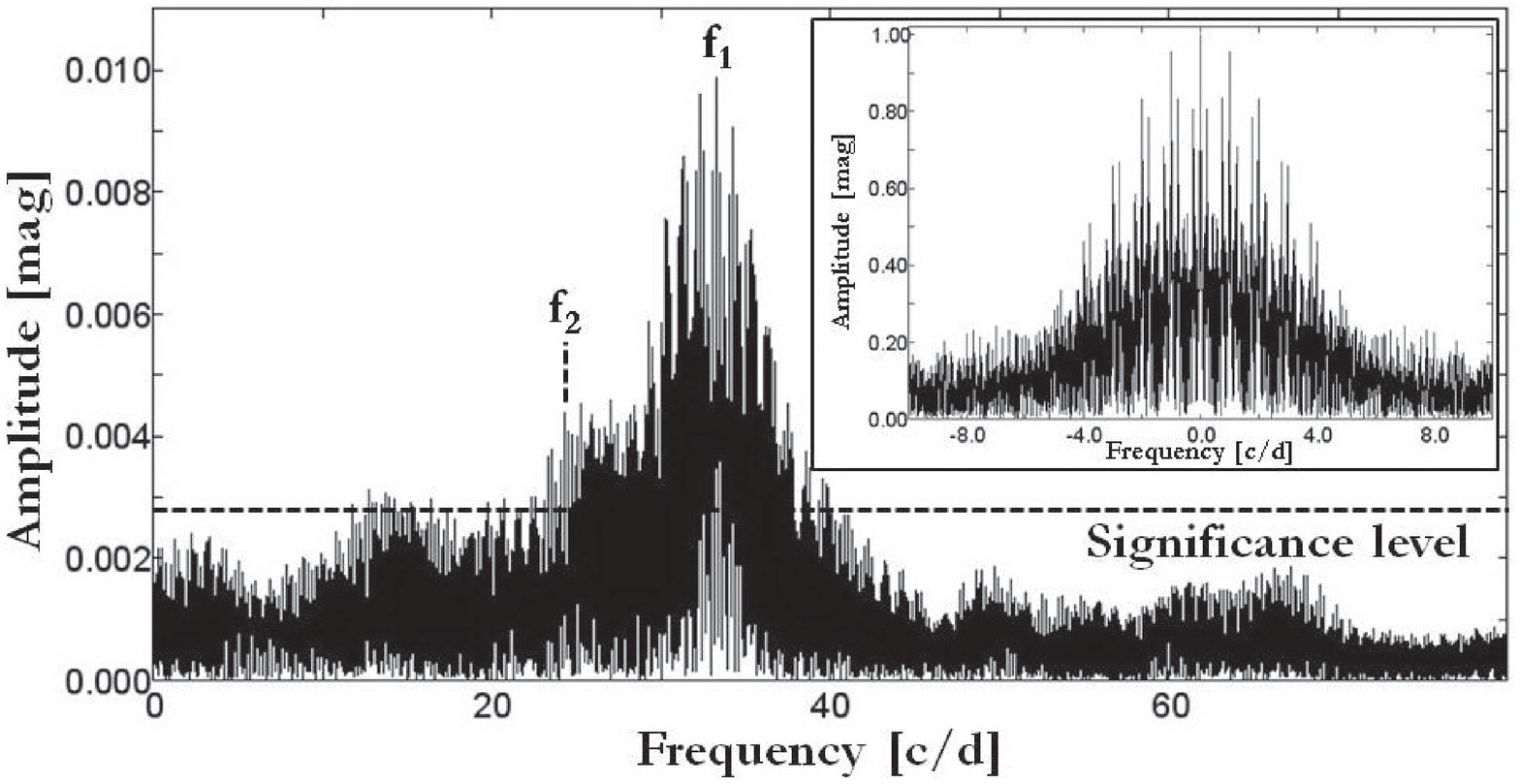}&\includegraphics[width=7.5cm]{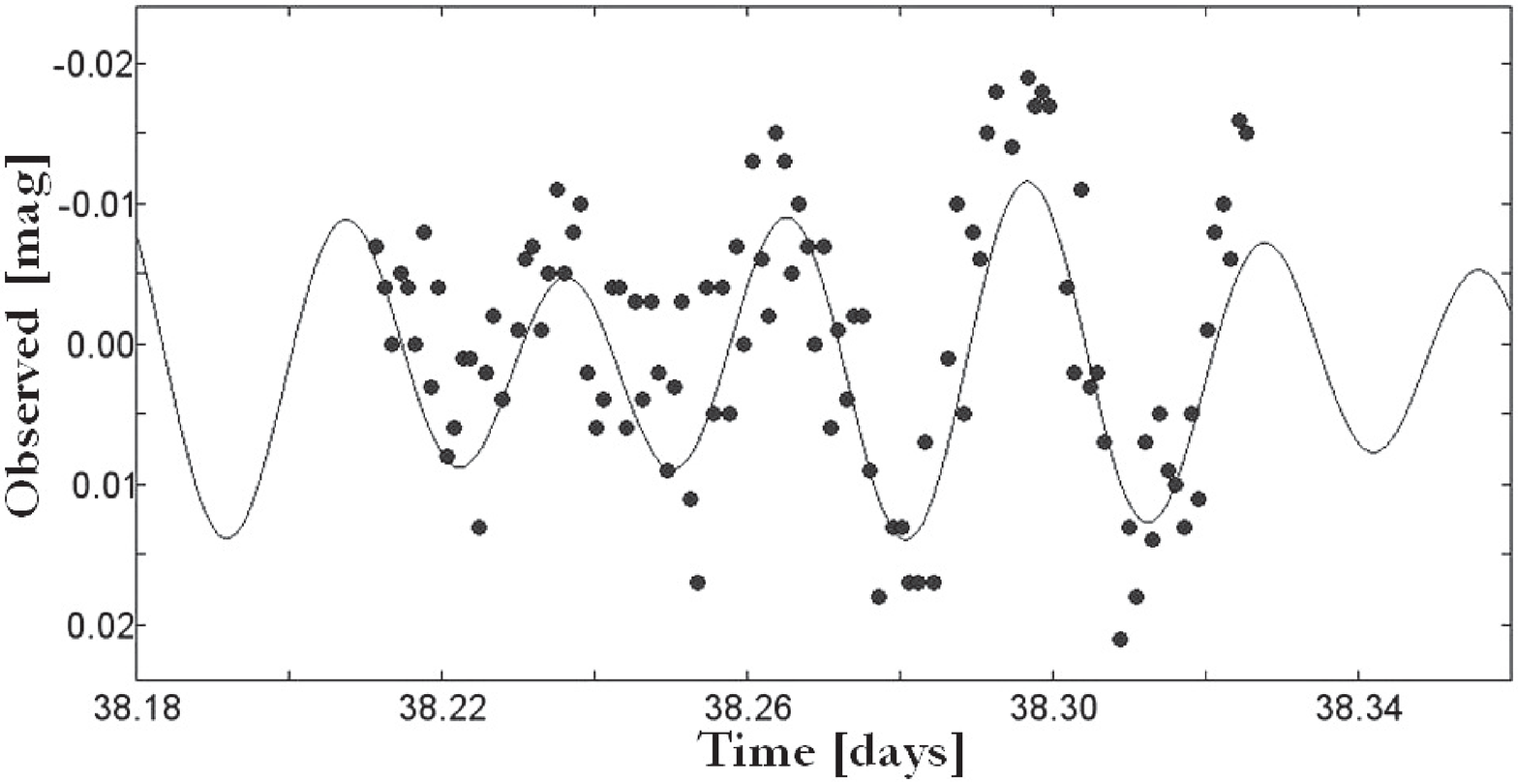}&(e)\\
\end{tabular}
\label{fig4}
\caption{Amplitude spectra (left panels) where the detected frequencies, the significance level (4$\sigma$) and spectral window plots (internal panels) are indicated, and Fourier fits on the longest data sets (right panels) for: (a) QY~Aql, (b) BW~Del, (c) TZ~Dra, (d) BO~Her and (e) RR~Lep.}
\end{figure*}

\section{Orbital period analysis}
\label{O-C}

%%Intro
Since all systems, except for RR~Lep, show interesting period changes, O$-$C diagram analysis was performed in order to find the mechanisms forming their orbital periods. TZ~Dra and BO~Her show cyclic period modulations, therefore the $LIght-Time~Effect$ (hereafter LITE) \citep{WO22,IR59} and the Applegate's  mechanism \citep{AP92} were tested. On the other hand, QY~Aql and BW~Del present secular period changes, so the mechanisms of mass transfer and mass loss due to possible magnetic braking effect were examined for implications in their orbital period.

%%LITE
Computation of the LITE parameters is a classical inverse problem for several derivable parameters; namely, period $P_{3}$ and eccentricity $e_3$ of the third body's wide orbit, HJD of the periastron passage $T_0$, semi-amplitude $A$ of the LITE and argument of periastron $\omega_3$. The ephemeris parameters ($JD_0$ and $P$ for the linear form and $C_{2}$ for the quadratic) were calculated together with those of the LITE. The LITE mass function $f(M_3)$ \citep[cf.][]{LQ09}:
\begin{equation}
f(M_3) = \frac{1}{P_3^2}\left[\frac{173.145 A}{\sqrt{1 - e_3^2 \cos^2 \omega_3}} \right]^3 = \frac{(M_3 \sin i_3)^3}{(M_1+M_2+ M_3)^2}
\end{equation}
with the wide orbit's period $P_3$ in yr, and the LITE amplitude $A$ in days, therefore produces the minimal mass of the tertiary component $M_{3\mathrm{,min}}=M_3 \sin i_3$ (with $i_3=90^\circ$). 
%%Applegate's mechanism
Late type components of EBs can be expected to present magnetic activity. The observed cyclic period changes may therefore come from variation of the magnetic quadrupole moment $\Delta Q$ \citep{AP92}.  \citet{AP87} and \citet{RO00} suggested the following formulae, respectively, for the $\Delta Q$ calculation:
\begin{eqnarray}
\frac{\Delta P}{P} &=& -9 \frac{\Delta Q}{Ma^2}\,\,   ,\\
%&&\nonumber \\
\Delta P &=& A \sqrt{2 [1 - \cos (2\pi P/P_{3})]}\,\,   ,
\end{eqnarray}
where $P$ and $a$ are the binary's period and semi-major axis, respectively, $P_3$ and $A$ the period and the semi-amplitude of the variation, respectively, and $M$ the mass of the potential magnetically active star (i.e. the secondary components of the studied cases). According to \citet{LR02}, magnetic activity results in detectable period modulation when $\Delta Q$ ranges between $10^{50}-10^{51}$~g~cm$^2$.

%%Secular changes
Mass transfer as well as mass and angular momentum loss due to magnetic braking are mechanisms that produce secular orbital period changes \citep{HI01}. The O$-$C analysis derives the quadratic term $C_2$ that can be used to calculate the orbital period change rate $\dot{P}$. Using the derived $\dot{P}$ and the parameters of the system's components, the mass transfer $\dot{M}_{\rm tr}$ ($>0$ for classical Algols) and the mass loss $\dot{M}_{\rm loss}$ ($<0$) rates can follow using the following formulae of \citet{HI01} (i.e. for conservative mass transfer) and \citet{ER05} (i.e. for mass loss due to magnetic braking and mass transfer between the components), respectively:
\begin{eqnarray}
\dot{M_{\rm tr}} &=& \frac{\dot{P}}{3P} \frac{M_1~M_2}{M_1-M_2} \,\,\,   ,\\
%&&\nonumber \\
\frac{\dot{P}}{3P}&=&k^2 \left( \frac{r_{\rm A}}{a} \right)^2 \frac{M_1+M_2}{M_1~M_2} \dot{M}_{\rm loss}+ \frac{M_2-M_1}{M_1~M_2} \dot{M}_{\rm tr}\,\,   ,
\end{eqnarray}
where $k$ is the gyration constant of the mass looser, $r_{\rm A}$ is the Alfv\'{e}n radius, $a$ is the semi-major axis of the system's orbit, and $M_1,~M_2$ the masses of the components.

%%%%%%%%%%%%%%%%%%%%%%%%%%%%%%%%%%%%%%%%%%%%%%%%%%%%%%%%%%%%%%%%%%%%%%%%%%%%%%%%%%%%%%%%%%%%%%%%%%%%%%TABLE 5----------- O-C
\begin{table}[h]
\caption{O$-$C diagram analyses results. The errors are indicated in parentheses alongside adopted values.}
\label{tab5}
\centering
\scalebox{0.92}{
\begin{tabular}{l cc}
\tableline 				
Parameters	                        &	   QY Aql	&	           BW Del	         \\
\tableline					
	                                &\multicolumn{2}{c}{\textsl{Eclipsing binary}}	 \\
\tableline					
$JD_0$~(HJD-2400000)	            &  37453.205 (1)&	       37375.460 (1)	     \\
$P$ (d)                             &	7.229560 (1)&	        2.423133 (3)	     \\
$C_{2}~(\times10^{-10}$~d/cycle)	&	-191 (8)	&	          40 (5)	         \\
$\dot{P}~(\times10^{-7}$~d/yr)	    &	   -19 (1)	&	          12 (2)	         \\
$\dot{M}_{\rm tr}~(\times10^{-8}$~M$_{\odot}$/yr)&  1$^{\rm a}$ &	   5.0 (6)	     \\
$\dot{M}_{\rm loss}~(\times10^{-8}$~M$_{\odot}$/yr)&-3.2 (1) &	      -- 	         \\
\tableline	
	                                &	     TZ Dra	&	           BO Her	         \\
\tableline					
	                                &\multicolumn{2}{c}{\textsl{Eclipsing binary}}	 \\
\tableline					
$JD_0$~(HJD-2400000)	            &  33852.346 (2)&	41884.620 (2)	             \\
$P$ (d)                             &	0.866033 (1)&	4.272834 (2)	             \\
\tableline	
	                                &\multicolumn{2}{c}{\textsl{LITE and third body}}\\
\tableline					
$T_0$~(HJD-2400000)	                &  53466 (480)	&	51000 (1247)	             \\
$\omega_3~(^\circ$)	                &	167 (10)	&	76 (40)	                     \\
$A$~(d)  	                        &	0.012 (1)	&	0.028 (2)	                 \\
$P_3$~(yr)                          &	62 (3)	    &	31.3 (7)	                 \\
$e_3$	                            &	0.5 (1)	    &	0.2 (1)	                     \\
$f(M_3)$~(M$_{\odot}$)	            &	0.0036 (1)	&	0.111 (1)	                 \\
$M_{\mathrm{3,min}}$~(M$_{\odot}$)  &	0.29 (1)	&	1.06 (1)	                 \\
\tableline	
	                                &\multicolumn{2}{c}{\textsl{Quadrupole moment variation}}\\
\tableline					
$\Delta Q~(\times10^{50}$g cm$^2$)  &	1.1    &	     14	                     \\
\tableline	
\multicolumn{3}{l}{$^{\rm a}$assumed}
\end{tabular}}
\end{table}

%%Technique
62 times of minima for QY~Aql, 48 for BW Del, 171 for TZ~Dra, and 40 for BO~Her, taken from literature and minima databases\footnote{http://var.astro.cz/ocgate/}, were used for the O$-$C analyses. The systems' ephemerides of \citet{KR01} were used to compute, initially, the O$-$C points from all the compiled data. The analysis was based on least squares method with statistical weights on a \textsl{MATLAB} code \citep{ZA09}. Weights were set at $w=1$ for visual, 5 for photographic and 10 for CCD and photoelectric data. In Fig.~5 full circles represent times of primary minima and open circles those of the secondary minima, where the bigger the symbol, the bigger the weight assigned. The corresponding parameters of the solutions are listed in Table~\ref{tab5}.

%%adopted solutions & brief Results
The O$-$C points of TZ~Dra and BO~Her show cyclic distribution, therefore the LITE and the Applegate's mechanisms were tested by fitting the respective periodic curves. Moreover, a parabolic term, in accordance with the potential mass transfer from the secondary to the primary (i.e. conventional semidetached configurations; see Section~\ref{DA}), was also tested in the fittings, but it resulted in unrealistic values, hence we excluded it. BO~Her was found to have $\Delta Q$ value marginally outside the range that can produce cyclic period changes \citep{LR02}, therefore, the LITE seems to be the most possible explanation for its orbital period modulations. On the contrary, both the LITE and the Applegate's mechanism can explain the cyclic period changes of TZ~Dra.

%%%%%%%%%%%%%%%%%%%%%%%%%%%%%%%%%%%%%%%%%%%%%%%%%%%%%%%%%%%%%%%%%%%%%%%%%%%%%%%%%%%%%%%%%%%%%%%%%%%%%% Figure 4 -- O-Cs
\begin{figure}[h!]
\centering
\begin{tabular}{cl}
%QY Aql\\
\includegraphics[width=7.1cm]{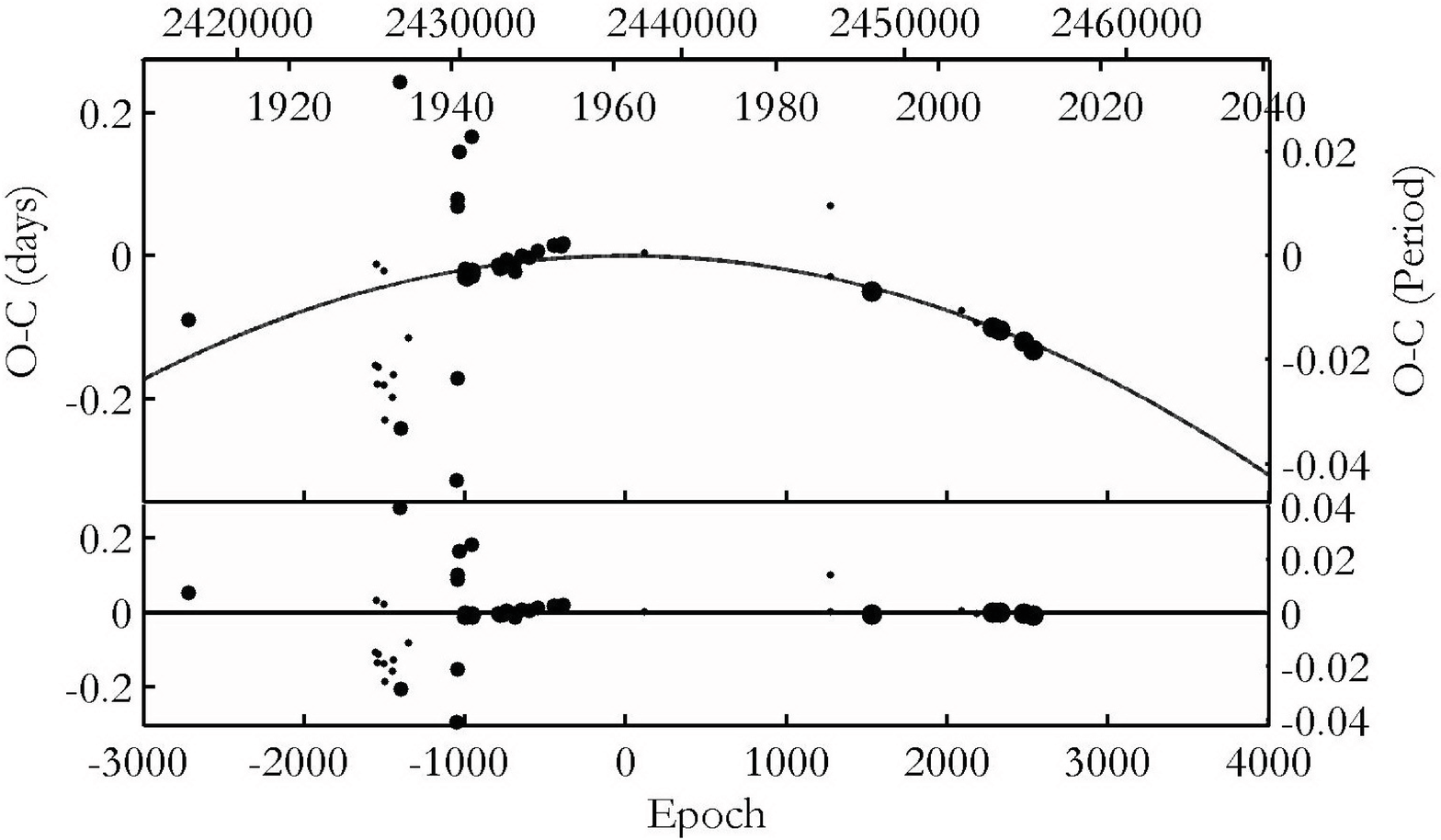}&(a)\\
%BW Del                                     \\
\includegraphics[width=7.1cm]{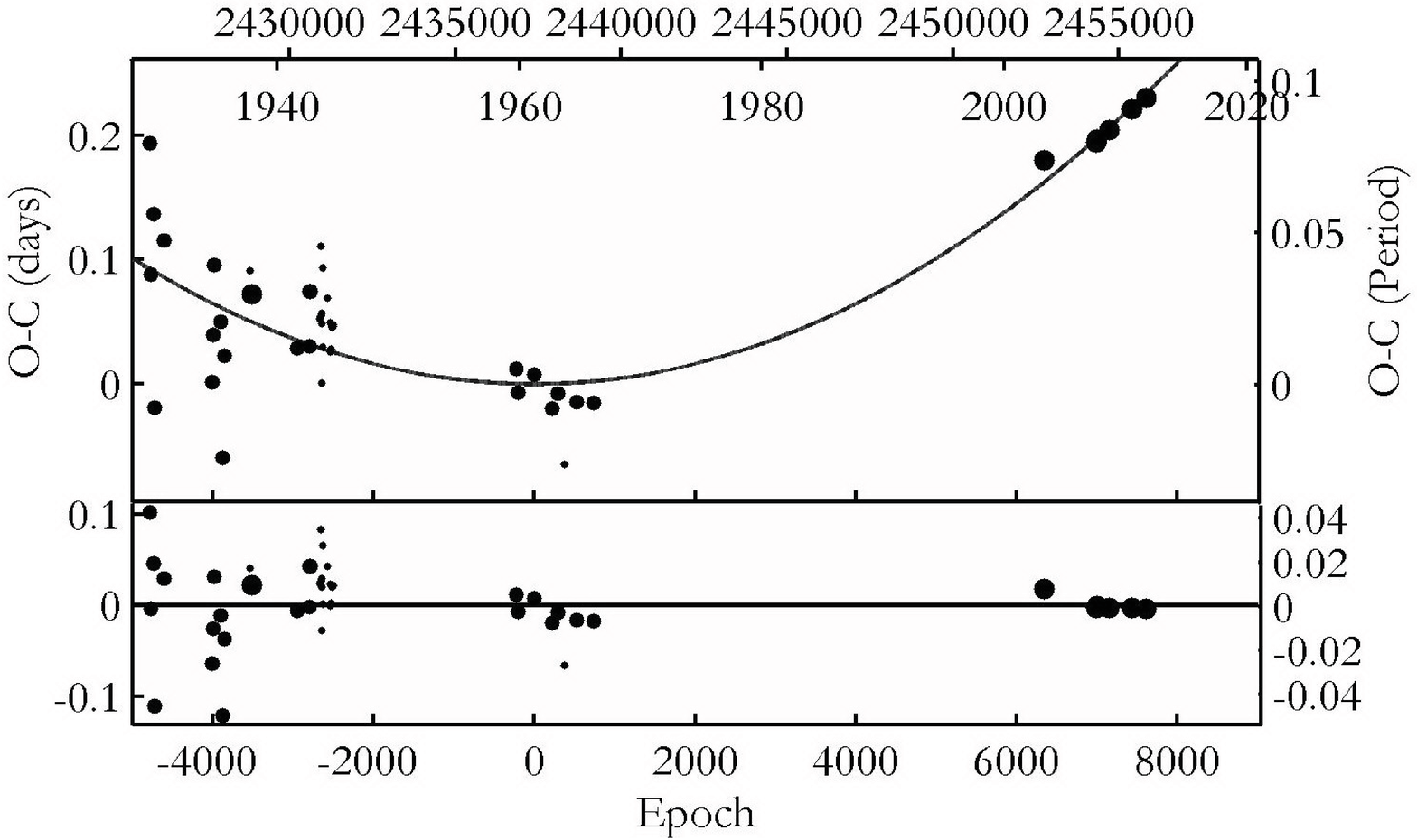}&(b)\\
%TZ Dra                                     \\
\includegraphics[width=7.1cm]{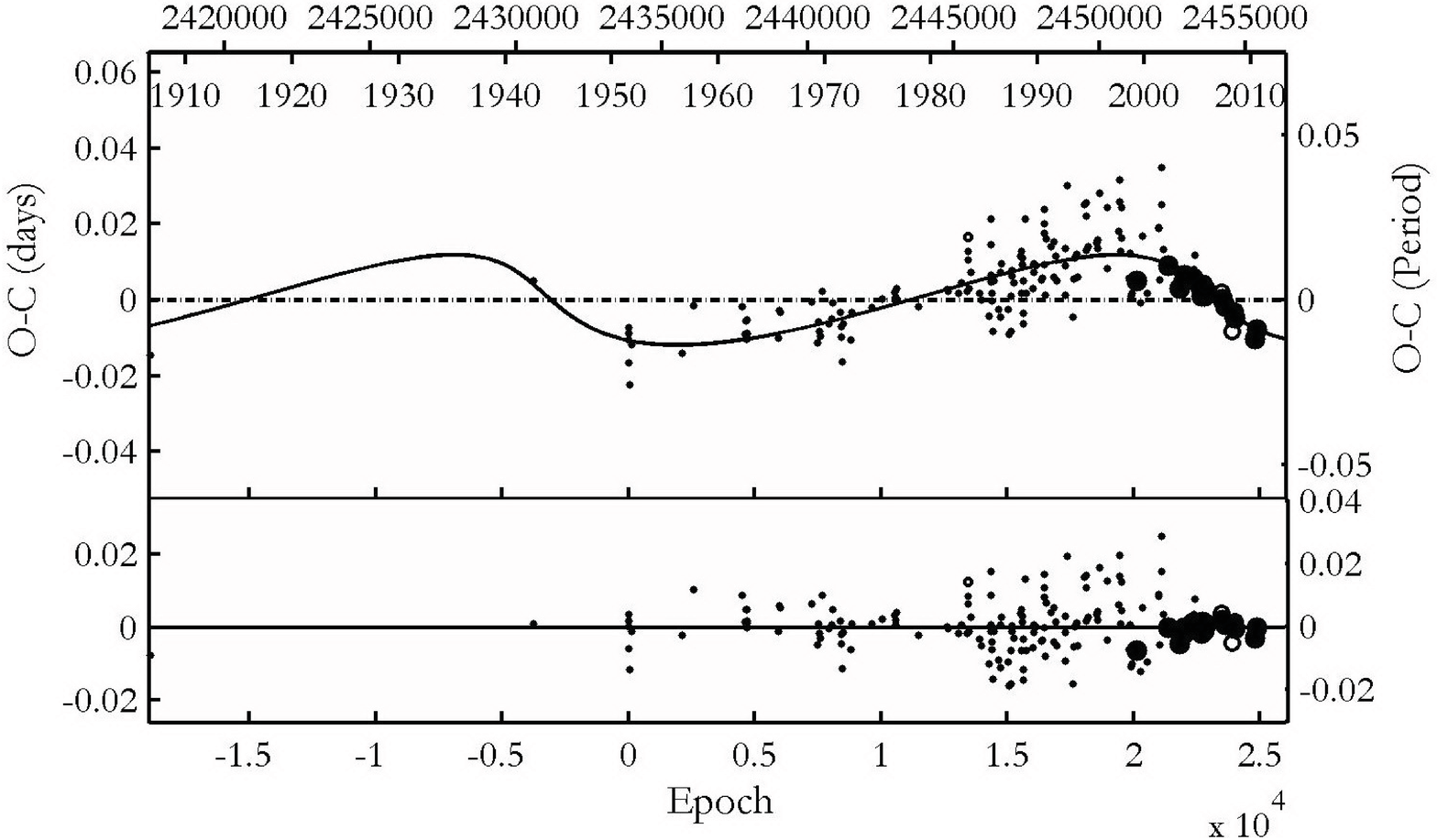}&(c)\\
%BO Her\\
\includegraphics[width=7.1cm]{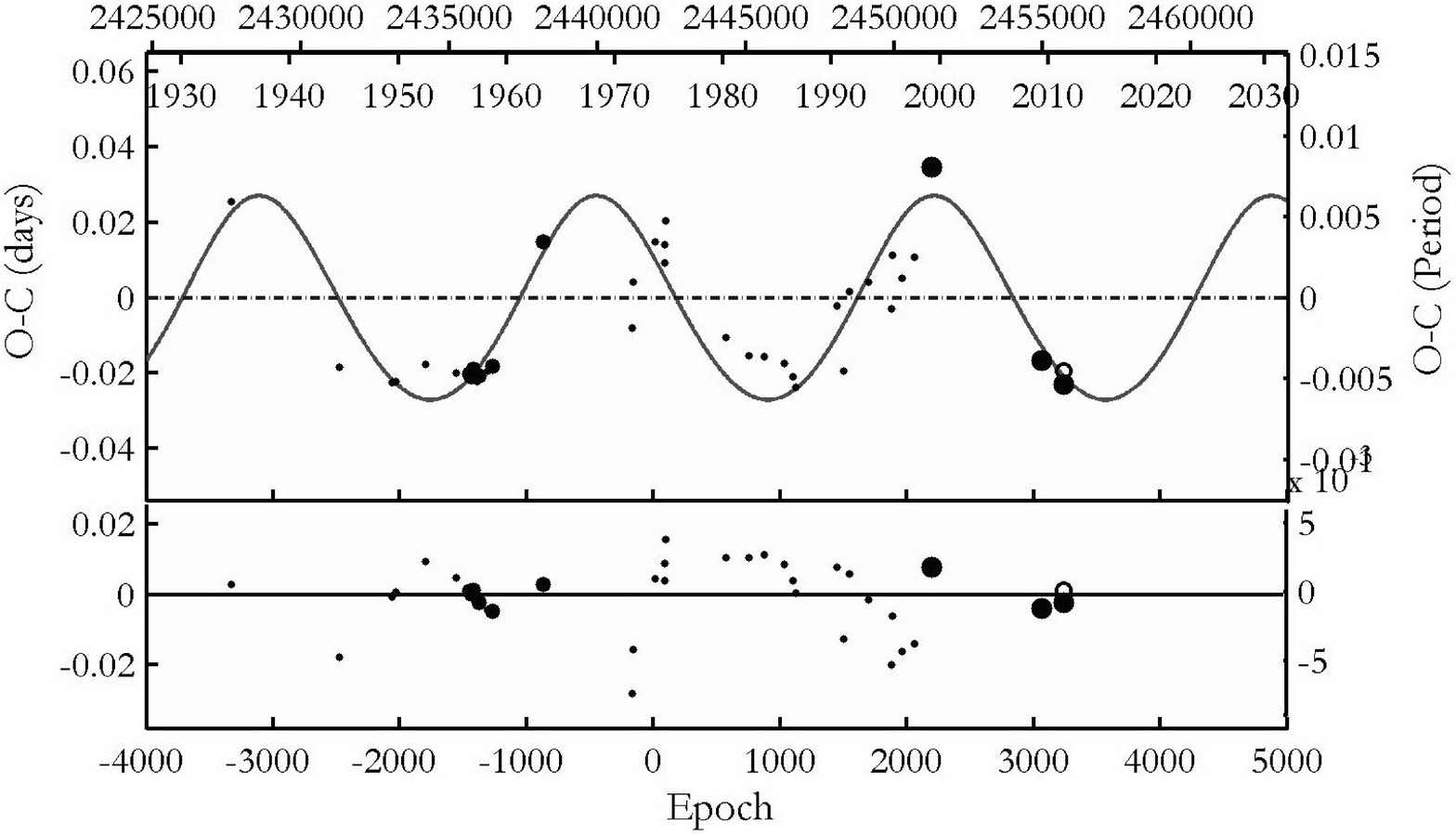}&(d)\\
\end{tabular}
\caption{O$-$C diagrams of all systems fitted by theoretical curves (upper part) and the residuals after the subtraction of the adopted solution (lower part) for (a) QY~Aql, (b) BW~Del, (c) TZ~Dra and (d) BO~Her.}
\label{fig5}
\end{figure}

For BW~Del and QY~Aql a parabola was chosen for fitting their O$-$C points, since mass flow from the secondaries to the primaries is expected to occur in accordance with the secondaries' Roche lobe filling (see Section~\ref{DA}). BW~Del indeed shows a secular period increase due to mass transfer from its less to its more massive component. Although we expected the same for QY~Aql, it was found that its period decreases with a rapid rate. However, the observed period changes can be interpreted with a combination of two mechanisms, namely the magnetic braking effect of the secondary component, which causes mass loss from the system, and the mass transfer from the secondary to the primary. A similar case (BG Peg) regarding the geometrical configuration and the period decrease was studied by \citet{SO11}, so the same values (i.e. $\dot{M}_{\rm tr}\sim10^{-8}~$M$_{\odot}$/yr, $k^2\cong0.1$, and $r_{\rm A}=10R_2$) were assumed for a rough estimate of the mass loss rate of QY~Aql due to magnetic braking.

The system RR~Lep has a rather steady constant orbital period, therefore its O$-$C diagram was not analysed. However, it is available on the O$-$C Atlas of \citet{KR01}.

\section{Discussion and conclusions}
\label{DIS}

%%%%Overview
One newly discovered (BW~Del), and four already known (QY~Aql, TZ~Dra, BO~Her and RR~Lep) eclipsing systems with a $\delta$~Sct component were observed and analysed using modern analysis tools in order to obtain useful conclusions about their oscillating behaviour, geometrical shape, absolute parameters, evolutionary stage and orbital period modulations. These systems are confirmed as classical Algols with their primaries showing $\delta$~Sct type pulsations. Therefore, according to the definition given by \citet{MK04}, they can also be considered as oEA systems. LT~Her was also identified as an EB including a $\delta$~Sct type member, but the detailed results will be presented in the future. Four other EBs, candidates for including $\delta$~Sct components, namely V345~Cyg, MX~Her, TW~Lac and AQ~Tau, were also checked for pulsations but the results were negative.

%% QY Aql
The primary component of QY~Aql is located beyond the TAMS and pulsates with a frequency of $\sim10.656$~c/d. The decreasing orbital period rate of the system is well explained with the mass transfer process from the secondary to the primary component, which is supported by its conventional semi-detached geometrical status, and the mass loss due to magnetic braking of its secondary, which was found to be at the giant stage of evolution. A mass loss rate of $3.2\times10^{-8}$~M$_{\odot}$/yr, typical for red giants \citep{HI01}, was estimated.

%% BW Del
Three pulsational frequencies were detected for the primary of BW~Del, with the most dominant one at 25.1~c/d. Based on the adopted mass and the derived radius, the star is located beyond the TAMS, but very close to it. The secondary component was found to be very evolved and it transfers material to the primary with a rate of $5\times10^{-8}$~M$_{\odot}$/yr.

%%TZ Dra
The primary component of TZ~Dra is a relatively fast pulsator with a frequency of $\sim50.99$~c/d and is located near the ZAMS. The frequency analysis results are in agreement with those of \citet{MK05}. Based on its frequency value and its evolutionary status we conclude that its oscillating lifetime must have started recently, according to the evolutionary stage-pulsation period empirical relation for this kind of stars \citep{LI12}. The secondary component of the system has filled its Roche lobe and is located beyond the TAMS. The cyclic changes of the system's period are caused probably due to a tertiary component with a period of $\sim62$~yr and a minimal mass of $\sim0.3~M_{\odot}$. On the other hand, the LC analysis did not reveal any third light. However, assuming that the third body is a MS star, and based to the mass-luminosity relation for dwarfs ($L\sim M^{3.5}$), we can calculate its luminosity and compare it with the absolute luminosity values of the binary's members (see Table~\ref{tab3}) by using the following formula:
\begin{equation}
L_{3,{\rm O-C}} (\%)=100 \frac{M_{3,{\rm min}}^{3.5}}{L_1+L_2+M_{3,{\rm min}}^{3.5}}
\end{equation}
We found that the expected luminosity contribution of such a third star should be $\sim$0.14\%, hence its light absence is plausible. However, according to the value of $\Delta Q$ ($\sim10^{50}~$g~cm$^2$), it is possible that the period changes can be caused due to magnetic influence of the secondary component. Applegate's mechanism predicts also brightness changes of the system, but this has not been verified so far. Therefore, future photometric observations covering several decades and/or astrometric observations are needed in order to conclude about the mechanism that forms the binary's orbital period.

%%BO Her
For the oscillating member of BO~Her we traced two pulsation frequencies with the dominant one at 13.43~c/d. This results agrees with that of \citet{SB07}. Due to the relatively high amplitude of this mode, its first two harmonics of its pulsation frequency were also detected in the frequency spectrum. The primary (pulsating) component of the system is located on the TAMS edge. On the other hand, the secondary is located far beyond the TAMS, being at the giant stage of evolution. A third body with a minimal mass of $\sim1.1~M_{\odot}$ and a period of $\sim31$~yr may exist around the EB, but we did not detect any additional luminosity in the LC analysis. Following the same method as for the case of TZ~Dra, we found an expected luminosity contribution $\sim$5\%, which is large enough to be detected photometrically. The most possible explanation for this disagreement could be either the non-MS nature of the third body (e.g. exotic object) or that the third body is in fact a binary with two low-mass and low-temperature components, providing lower luminosity in total, instead of a single star. Future spectroscopic and/or astrometric observations are desirable in order to solve this mystery.

%% RR Lep
The primary component of RR~Lep pulsates in two modes with the dominant frequency at $\sim33.28$~c/d and it is located on the MS and very close to the TAMS. The present results regarding the frequency $f_1$ are in marginal agreement with those of \citet{DV09}, who found only one pulsation frequency of $\sim31.87$~c/d. However, our results are based on two-filter data which were obtained with better equipment in comparison with that used by \citet{DV09}. The secondary of the system is a rather more evolved star located above the TAMS.

%% Mass transfer
The O$-$C points distributions of TZ~Dra, BO~Her and RR~Lep do not show any secular period changes that can be connected with mass transfer. Very probably, these systems are at slow mass-accretion stage \citep{MK03} with a rate that cannot be detected with the current time coverage of minima timings.

%%%Comparison with models
\citet{LI12}, based on the pulsational and absolute parameters of the $\delta$~Sct components of all known oEA stars, derived empirical relations between the dominant pulsation period $P_{\rm puls}$ and the gravity acceleration value $g$ and between the $P_{\rm puls}$ and the orbital period $P_{\rm orb}$ of the systems. Therefore, it is useful to check if the respective values of the pulsating components of the systems analysed herein follow these trends, and this is shown in Fig.~6.

%%%%%%%%%%%%%%%%%%%%%%%%%%%%%%%%%%%%%%%%%%%%%%%%%%%%%%%%%%%%%%%%%%%%%%%%%%%%%%%%%%%%%%%%%%%%%%%%%%%%%% Figure 5 -- comparison with models
\begin{figure}[h]
\centering
\begin{tabular}{c}
\includegraphics[width=7.8cm]{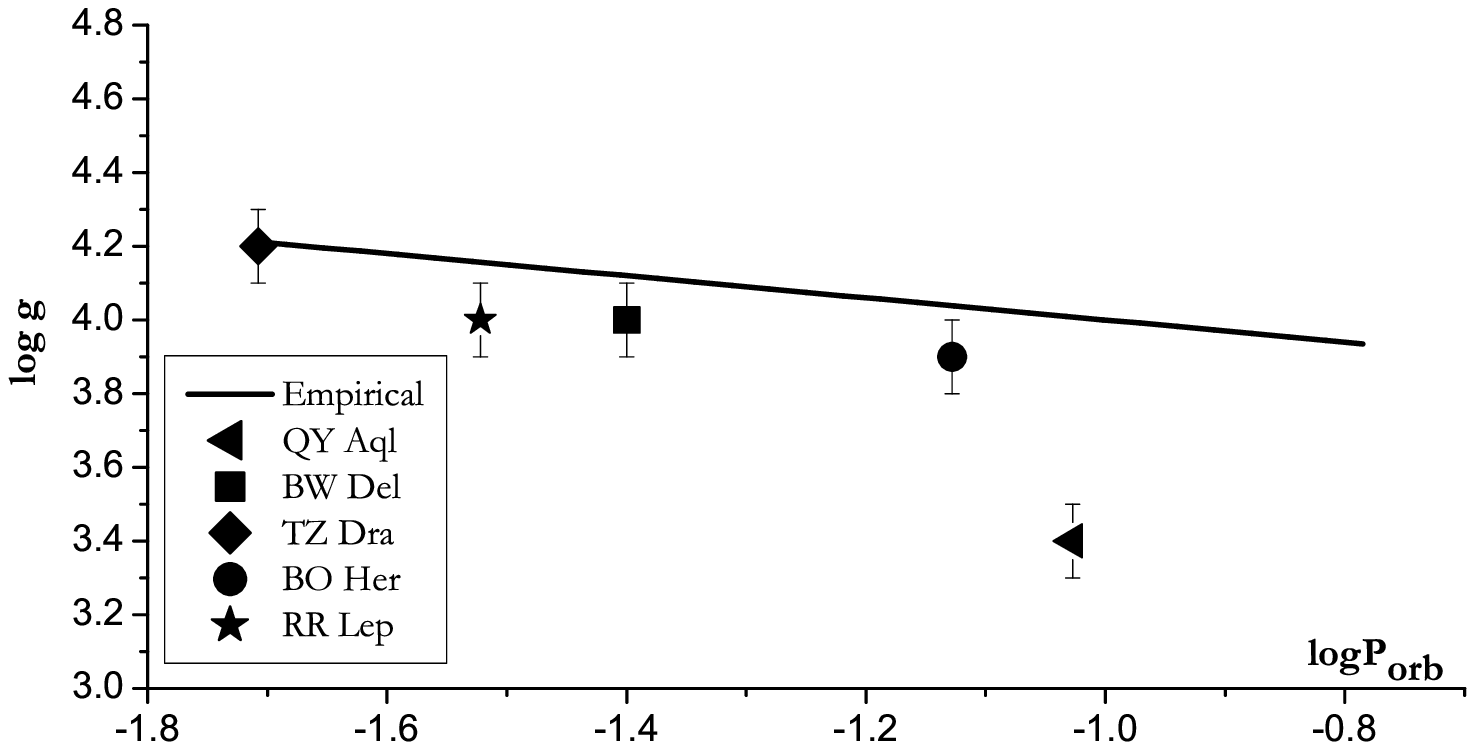}\\
\includegraphics[width=7.8cm]{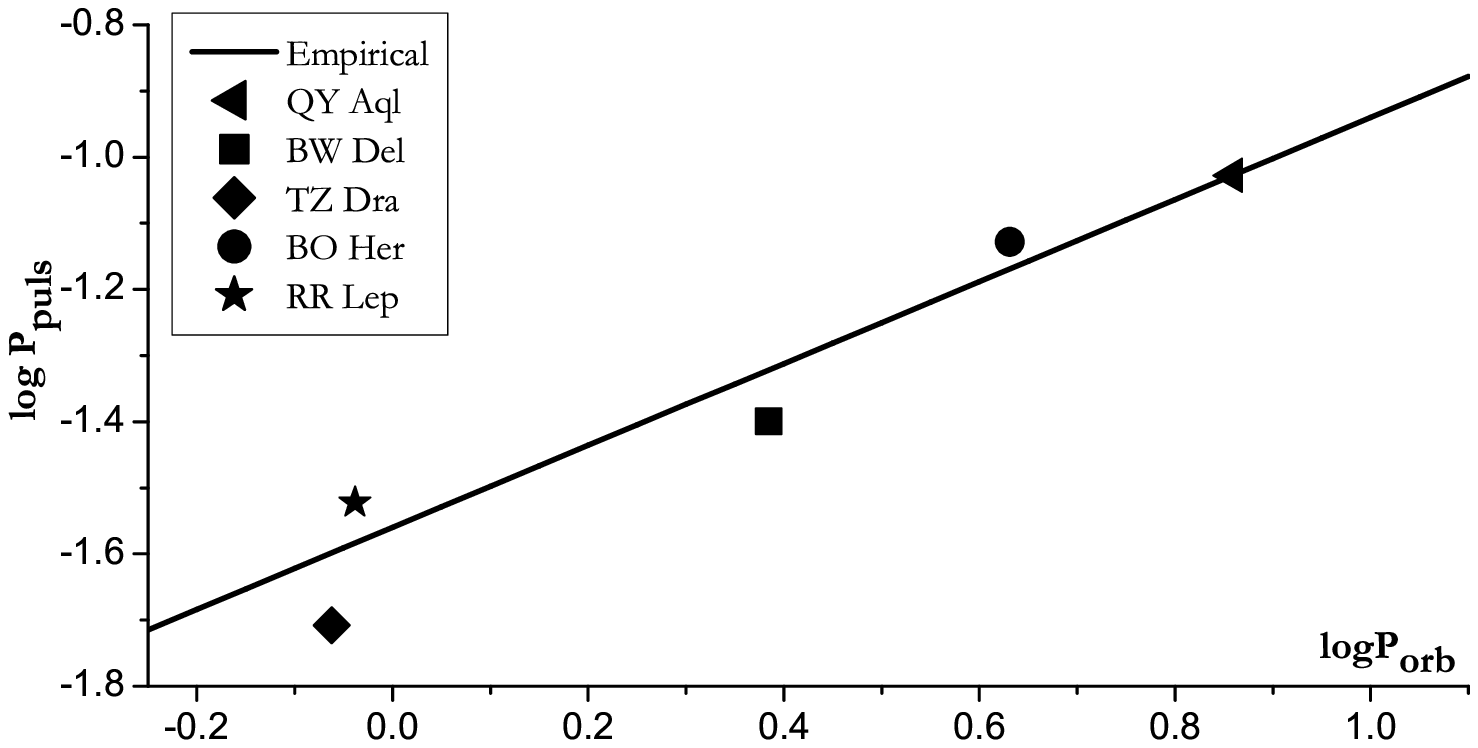}\\
\end{tabular}
\label{fig6}
\caption{Positions of the pulsating members of the systems in the $g-P_{\rm puls}$ (upper) and $P_{\rm puls}-P_{\rm orb}$ (lower) diagrams for oEA stars with $\delta$~Sct members \citep{LI12}. The error bars of the periods' values are not shown due to scale reasons.}
\end{figure}

The pulsating stars of all systems seem to follow well the $P_{\rm puls}-P_{\rm orb}$ and $g-P_{\rm puls}$ trends, with the exception of the primary of QY~Aql in the $g-P_{\rm puls}$ diagram. This star is at the subgiant evolutionary stage. On the other hand, the sample, in which the empirical relation of $g-P_{\rm puls}$ is based, consists mostly of MS stars. Therefore, QY~Aql, with the longest orbital period and minimum $g$-value in the sample of \citet{LI12}, might have followed a different evolutionary track or another relation between $g-P_{\rm puls}$ for the evolved oEA stars has to be examined.

%% future work and suggestions
Radial velocities measurements for all systems are needed in order to determine their absolute parameters and the $l$-degrees of their pulsation modes with higher certainty. Moreover, more precise photometric observations (e.g. space data) are expected to reveal additional pulsation frequencies that they could not be detected with the present instrumentation setup. Future surveys aiming to new discoveries of this kind of systems and long-term monitoring of the already known ones are highly encouraged in order to enrich our knowledge about the mass transfer implication in the pulsation mechanisms and, in general, about the stellar evolution of binaries with A-F components.

\section*{Acknowledgments}
This work is part of the Ph.D thesis of A.L. and has been financially supported by the Special Account for Research Grants No 70/4/11112 of the National \& Kapodistrian University of Athens, Hellas. In the present work, the minima database: (http://var.astro.cz/ocgate/), SIMBAD database, operated at CDS, Strasbourg, France, and Astrophysics Data System Bibliographic Services (NASA) have been used.

\end{document}